%% file: main.tex
\documentclass{IEEEtaes_nofooter}

\input{preamble}

\usepackage[citestyle=numeric-comp,bibstyle=ieee]{biblatex}
\AtBeginBibliography{\footnotesize}

\newcommand{\revised}[1]{#1}
\newenvironment{revisedblock}{}{\ignorespacesafterend}

\jvol{XX}
\jnum{XX}
\jmonth{XXXXX}
\paper{1234567}
\pubyear{YYYY}
\doiinfo{TAES.YYYY.Doi Number}

\begin{document}

\title{Efficient Localization of Directional Emitters via Joint Beampattern Estimation}

\author{FRASER WILLIAMS\orcidlink{0000-0002-5422-2827}}
\affil{Queensland University of Technology, Brisbane, Australia}
\author{AKILA PEMASIRI\orcidlink{0009-0003-2694-7534}}
\affil{Queensland University of Technology, Brisbane, Australia}
\author{DHAMMIKA JAYALATH\orcidlink{0000-0001-6130-0275}}
\member{Senior Member, IEEE}
\affil{Queensland University of Technology, Brisbane, Australia} 
\author{TERRENCE MARTIN\orcidlink{0000-0002-2166-9796}}
\affil{Revolution Aerospace Pty Ltd, Brisbane, Australia} 
\author{CLINTON FOOKES\orcidlink{0000-0002-8515-6324}}
\member{Senior Member, IEEE}
\affil{Queensland University of Technology, Brisbane, Australia}

\receiveddate{The research for this paper received funding support from Revolution Aerospace Pty Ltd and the Queensland Government through Trusted Autonomous Systems (TAS), a Defence Cooperative Research Centre funded through the Commonwealth Next Generation Technologies Fund and the Queensland Government.}

\corresp{{\itshape (Corresponding author: F. Williams).}}

\authoraddress{Authors' addresses: F. Williams, A. Pemasiri, D. Jayalath, and C. Fookes are with the School of Electrical Engineering and Robotics, Queensland University of Technology, Brisbane, QLD 4000, Australia (e-mail: \href{mailto:f6.williams@qut.edu.au}{f6.williams@qut.edu.au}; \href{mailto:a.thondilege@qut.edu.au}{a.thondilege@qut.edu.au}; \href{mailto:dhammika.jayalath@qut.edu.au}{dhammika.jayalath@qut.edu.au}; \href{mailto:c.fookes@qut.edu.au}{c.fookes@qut.edu.au}). T. Martin is with Revolution Aerospace Pty Ltd, Brisbane, QLD, Australia (e-mail: \href{mailto:terry@revn.aero}{terry@revn.aero}).}

\markboth{WILLIAMS ET AL.}{EFFICIENT LOCALIZATION OF DIRECTIONAL EMITTERS}
\maketitle

\begin{abstract}
Localization methods designed for omnidirectional emitters degrade on directional sources, where directive beampatterns induce pronounced received signal strength (RSS) modulations. We present a direct position determination (DPD) approach that augments standard angle of arrival (AOA) and time difference of arrival (TDOA) measurements with RSS modulation to jointly estimate emitter position and beampattern parameters. Two algorithms reduce the four-dimensional search to two dimensions: a block coordinate ascent (BCA) method that alternates between position and beampattern subproblems, and a profile likelihood method that concentrates the beampattern parameters out via a generalized eigenvalue solve, a weighted least squares fit, and Newton refinement on the constrained Rayleigh quotient. We derive the Cram\'er--Rao lower bound (CRLB) and benchmark against an AOA--TDOA baseline and a prior-art approach that assumes known beamwidth and estimates only the orientation. The profile likelihood method approaches the CRLB, substantially outperforms all baselines at low signal-to-noise ratio (SNR), and remains robust under model mismatch with the assumed von Mises template. A sensitivity analysis shows performance peaks when receivers lie on the main-lobe edges where RSS gradients are maximized.
\end{abstract}

\begin{IEEEkeywords}
Array processing, direct position determination, emitter localization, received signal strength, directional emitter
\end{IEEEkeywords}

\section{Introduction} \label{sec:intro}
Passive localization of radio frequency (RF) emitters is a fundamental capability across navigation \cite{hoSolutionPerformanceAnalysis1993}, telecommunications, and surveillance \cite{ketabalianClosedFormSolutionLocalization2020}. Estimators typically fuse measurements from multiple spatially distributed receivers using an a priori model of signal transformations \cite{weissDirectPositionDetermination2004}, including time difference of arrival (TDOA) \cite{hoSolutionPerformanceAnalysis1993}, angle of arrival (AOA) \cite{weissDirectPositionDetermination2004, liuJointTDOAAOA2013}, frequency difference of arrival (FDOA) \cite{musickiGeolocationUsingTDOA2008, songFastAlgorithmDirect2019}, and received signal strength (RSS) \cite{ketabalianClosedFormSolutionLocalization2020, youDirectPositionDetermination2023}.

Most localization estimators implicitly assume an omnidirectional emitter. Modern radio systems, however, routinely employ directive antennas or phased arrays for long-range communications, spatial multiplexing, and radar \cite{skolnikIntroductionRadarSystems1980, sibleyModernTelecommunicationsBasic2018}, producing pronounced RSS variations across spatially distributed receivers. Ignoring these variations introduces substantial localization error \cite{youJointDirectEstimation2023}. Few prior works address directional emitters explicitly: the most directly comparable, \cite{youJointDirectEstimation2023}, assumes the emitter beamwidth is known a priori and estimates only the orientation. As we show, localization performance is highly sensitive to beamwidth misspecification, motivating an estimator that recovers orientation and beamwidth jointly with position. \revised{To our knowledge, the present work is the first DPD method to estimate all three blindly, with no prior on the beamwidth.}

Direct position determination (DPD) incorporates signal measurements directly into a physics-based model via maximum-likelihood estimation (MLE), and has improved localization at low signal to noise ratio (SNR) by folding path loss into the transformation model \cite{williamsEnhancingEmitterLocalization2023}. \revised{This and related RSS and path-loss-aware DPD methods \cite{ketabalianClosedFormSolutionLocalization2020, youDirectPositionDetermination2023, zhaoContextualDirectPosition2025, youDirectPositionDetermination2026} model the isotropic link budget but assume an omnidirectional emitter.} We extend the DPD framework to directional emission by jointly estimating the emitter's position and beampattern parameters. Since emitters can employ arbitrary beampatterns, we introduce a generalized beampattern model that captures the main lobe through two parameters, orientation $\phi$ and half-power beamwidth $\beta$, at a fidelity sufficient for localization while remaining tractable.

Adding the beampattern parameters lifts the search from two dimensions (position) to four (position plus orientation and beamwidth), so that the cost of an exhaustive search grows in proportion to the beampattern grid resolution. To address this, we develop two algorithms that both reduce the four-dimensional joint search to two dimensions. The first is a block coordinate ascent (BCA) method that alternates between position and beampattern subproblems. The second is a profile likelihood method that analytically concentrates the beampattern parameters out: the inner solve recovers the unconstrained amplitude pattern via a generalized eigenvalue decomposition, fits the beampattern parameters by weighted least squares, and refines them with Newton iterations on the constrained Rayleigh quotient using the analytic Hessian.

We further derive the CRLB for position estimation under the generalized beampattern model. Beyond establishing a numerical performance floor, the bound depends explicitly on the RSS gradient at each receiver, with a concrete geometric implication: receivers contribute most when positioned on the steep flanks of the main lobe rather than at the beam peak or in the sidelobes.

The remainder of this paper is organized as follows. \Cref{sec:problem} formulates the directional emitter localization problem and introduces the generalized beampattern model. \Cref{sec:crlb} derives the CRLB for position estimation. \Cref{sec:dpdalg} presents the BCA and profile likelihood algorithms. \Cref{sec:sim} evaluates performance through Monte Carlo simulations against the AOA--TDOA baseline of \cite{weissDirectPositionDetermination2004} and the beampattern-aware approach of \cite{youJointDirectEstimation2023}. \Cref{sec:conclusion} discusses the results and concludes the paper.

\textit{Notation:} Vectors and matrices are represented by bold lower-case and upper-case respectively. The symbols $\tilde{(\cdot)}$, $\norm{\cdot}$, $(\cdot)^\T$, $(\cdot)^*$, $(\cdot)^\H$, $(\cdot)^\dagger$, $\triangleq$, and $\mathbb{E}[\cdot]$ denote the discrete Fourier transform, $l^2$ norm, transpose, conjugate, Hermitian, Moore-Penrose pseudoinverse, definition, and expectation respectively.

\section{Problem Formulation} \label{sec:problem}
Consider a stationary directional emitter, and $L$ spatially distributed receivers each equipped with a uniform linear array (ULA) having $M$ elements. The signal received by the $l$th receiver is given by
\begin{align}
    \bm{r}_l(t) &= d_l(\bm{p}, \bm{\psi})  \bm{a}_l(\bm{p}) s(t - \tau_l(\bm{p}) - t_0) + \bm{n}_l(t), \label{eq:time_domain}
\end{align}
where $\bm{r}_l(t)$ is the $M \times 1$ received signal vector at time $t$, $\bm{p}=[x, y]^\T$ is the emitter position, $d_l(\bm{p}, \bm{\psi})$ is the directional path attenuation incorporating beampattern effects, $\bm{\psi}=[\phi, \beta]^\T$ is the beampattern parameter vector where $\phi$ is the emitter orientation and $\beta$ is the emitter main-lobe beamwidth in radians, $\bm{a}_l(\bm{p})$ is the receiver steering vector, $s(t-\tau_l(\bm{p})-t_0)$ is the signal waveform at time $t$ transmitted at $t_0$, delayed by $\tau_l(\bm{p})$, and $\bm{n}_l(t)$ represents noise at the $l$th receiver. The model assumes line-of-sight propagation where all position-dependent attenuation is captured by $d_l(\bm{p}, \bm{\psi})$.

The directional path attenuation decomposes into an isotropic path loss term and a directional beampattern gain:
\begin{align}
    d_l(\bm{p}, \bm{\psi}) &= \rho_l(\bm{p})\, g_l(\bm{p}, \bm{\psi}), \label{eq:d_def}
\end{align}
where $\rho_l(\bm{p})$ captures geometric spreading independent of beam direction and $g_l(\bm{p}, \bm{\psi})$ captures the angular dependence of the emitter's radiation pattern. This decomposition is the key distinction from omnidirectional models, where $g_l = 1$ for all receivers.

The isotropic path attenuation is
\begin{align}
    \rho_l(\bm{p}) &\triangleq \frac{\kappa_l}{\norm{\bm{p}-\bm{u}_l}}, \label{eq:rho_def} \\
    \kappa_l &\triangleq \sqrt{\frac{P_T G_0 G_R \lambda^2}{\left(4\pi \right)^2 L_l}},
\end{align}
where $\kappa_l$ combines all position-independent parameters: $P_T$ the transmit power, $G_0$ the peak gain of the emitter, $G_R$ the gain of an individual antenna element, $\lambda$ the radio wavelength, and $L_l$ system losses such as filter insertion loss. The $l$th receiver is at position $\bm{u}_l=[x_l, y_l]^\T$. It is possible to calibrate receivers such that $\kappa_l$ is constant across receivers through compensation of $L_l$.

The beampattern gain is approximated by a von Mises structure
\begin{align}
    g_l(\bm{p}, \bm{\psi}) &= e^{\alpha(\beta)\left[\cos\left(\theta_l^{(t)}(\bm{p})-\phi\right)-1\right]} \label{eq:g_def}\\
    \alpha(\beta) &\triangleq -\frac{\log(2)/2}{\cos(\beta/2)-1},
\end{align}
where $\beta$ is the half-power beamwidth of $g_l$ in radians, and $\theta_l^{(t)}(\bm{p})$ is the angle from the transmitter to the $l$th receiver. The transmit angle is given by
\begin{align}
    \theta_l^{(t)}(\bm{p}) &= \arctantwo \left(y_l-y, x_l-x \right)
\end{align}
where $\arctantwo$ is the quadrant-aware arc-tangent function and the $l$th receiver position is $\bm{u}_l=[x_l, y_l]^\T$. 

\begin{figure}[!t]
    \centering
    \includegraphics[width=7cm]{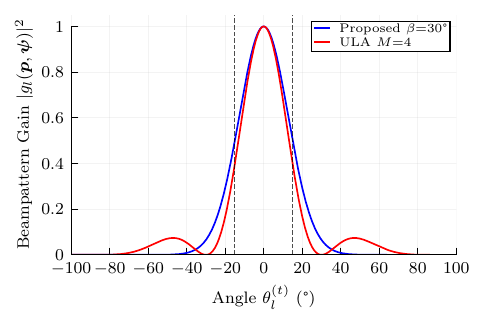}
    \caption{Generalized beampattern model ($\phi=0\degree$, $\beta=30\degree$) compared to a 4-element ULA of approximately $26\degree$ half-power beamwidth. Dashed lines mark the model's half-power beamwidth points.}
    \label{fig:generalised_beampattern_model}
\end{figure}

\revised{The von Mises form is smooth, even, and $2\pi$-periodic, and is chosen for tractability rather than physical fidelity. Its two parameters keep the joint search low-dimensional and enable the inexpensive inner solve of \cref{subsec:profile_likelihood}. Fig.~\ref{fig:generalised_beampattern_model} shows it agrees closely with a $4$-element ULA across the main lobe and diverges in the sidelobes. The template is a single symmetric lobe: it cannot reproduce sidelobes, nulls, or asymmetry, and as a real-valued gain it captures the RSS magnitude but not the emitter's angle-dependent phase response. \Cref{subsec:mismatch} quantifies this mismatch.}

The receiver steering vector is an $M \times 1$ vector with $m$th element
\begin{align}
    [\bm{a}_l(\bm{p})]_m &= e^{-j2\pi \frac{(m-1)\Delta}{\lambda} \cos(\theta^{(r)}_l(\bm{p}))}, \label{eq:steering_vector} \\
    \theta^{(r)}_l(\bm{p}) &= \arctantwo(y - y_l,x-x_l),
\end{align}
where $\Delta$ is the physical separation between antenna elements, $\lambda$ is the wavelength of the radio signal, and $\theta^{(r)}_l(\bm{p})$ is the angle from the $l$th receiver to the position $\bm{p}$.

The propagation delay is given by
\begin{align}
    \tau_l(\bm{p})&=\frac{\norm{\bm{p}-\bm{u}_l}}{c},
\end{align}
where $c$ is the speed of light in a vacuum.

Expressing (\ref{eq:time_domain}) in the frequency domain, we have
\begin{align}
    \tilde{\bm{r}}_{l,k} &= d_l(\bm{p}, \bm{\psi}) \bm{a}_l(\bm{p}) e^{-j\omega_k[\tau_l(\bm{p}) +t_0]} \tilde{s}_k + \tilde{\bm{n}}_{l,k}, \label{eq:freq_domain} \\
\omega_k&\triangleq \frac{2\pi k}{N_s T},
\end{align}
where $N_s$ is the number of signal samples, $T$ is the sampling period, $k$ is the discrete frequency bin, and $\tilde{s}_k$ and $\tilde{\bm{n}}_{l,k}$ are the Fourier coefficients of $s(t)$ and $\bm{n}(t)$ forming the vectors $\tilde{\bm{s}}=[\tilde{s}_0 \dots{} \tilde{s}_{N_s-1}]^\T$ and $\tilde{\bm{n}}=[\tilde{\bm{n}}_0 \dots{} \tilde{\bm{n}}_{N_s-1}]^\T$ respectively. Since the transmission time $t_0$ is common to all receivers, the factor $e^{-j\omega_k t_0}$ is absorbed into the signal coefficient $\tilde{s}_k \leftarrow e^{-j\omega_k t_0} \tilde{s}_k$.

Stacking all receivers, we define
\begin{align}
    \tilde{\bm{n}}_k &\triangleq [\tilde{\bm{n}}_{1,k}^\T \dots{} \tilde{\bm{n}}_{L,k}^\T]^\T \\
    \bm{a}_{k,l}(\bm{p}, \bm{\psi}) &\triangleq d_l(\bm{p}, \bm{\psi}) e^{-j\omega_k \tau_l(\bm{p})} \bm{a}_l(\bm{p}) \\
    \bm{a}_k(\bm{p}, \bm{\psi}) &\triangleq [[\bm{a}_{k,1}(\bm{p}, \bm{\psi})]^\T \dots{} [\bm{a}_{k,L}(\bm{p}, \bm{\psi})]^\T]^\T \label{eq:a_def} \\
    \tilde{\bm{r}}_k &\triangleq [\tilde{\bm{r}}_{1,k}^\T \dots{} \tilde{\bm{r}}_{L,k}^\T]^\T.
\end{align}
The received signal is then given by
\begin{align}
    \tilde{\bm{r}}_k &= \bm{a}_k(\bm{p}, \bm{\psi}) \tilde{s}_k + \tilde{\bm{n}}_k. \label{eq:model}
\end{align}

We wish to estimate the emitter position $\bm{p}$ using the observations of (\ref{eq:model}).

\section{Cramér-Rao Lower Bound} \label{sec:crlb}
We derive the Cramér-Rao Lower Bound (CRLB) for position estimation, which provides the minimum variance achievable by any unbiased estimator under our signal model.

\subsection{Fisher Information Matrix}
Noise is assumed to be drawn from the complex Gaussian distribution $\tilde{\bm{n}}_k \sim \mathcal{CN}(0, \sigma_n^2)$. The log-likelihood function of the observed signal is
\begin{align}
    \ln P(&\tilde{\bm{r}}) = -MLN_s\log(\pi\sigma_n^2) \nonumber \\
    &\qquad\ - \frac{1}{\sigma_n^2} \sum_{k=0}^{N_s-1}\norm{\tilde{\bm{r}}_k-\bm{a}_k(\bm{p}, \bm{\psi})\tilde{s}_k}^2 \label{eq:log-likelihood}.
\end{align}
The position bound is obtained from the Fisher information matrix (FIM) by concentrating out the nuisance parameters $\bm{\psi}$ and $\tilde{\bm{s}}$ via the partitioned-inverse identity.

The Fisher information matrix is given by
\begin{align}
    \bm{J} = -\mathbb{E} \left[\frac{\partial^2 \ln P(\tilde{\bm{r}})}{\partial \bm{\zeta}^2} \right] \label{eq:FIM_def}
\end{align}
where the vector $\bm{\zeta}$ contains all unknown parameters
\begin{align}
    \bm{\zeta} &\triangleq [\bm{p}, \bm{\epsilon}], \label{eq:zeta_def}\\
    \bm{\epsilon} &\triangleq [\bm{\psi}^\T, \tilde{\bm{s}}^\T]. \label{eq:epsilon_def}
\end{align}

The FIM has the partitioned structure
\begin{align}
    \bm{J}_{\bm{\zeta}\bm{\zeta}} &=%
        \frac{2}{\sigma_n^2}\begin{bmatrix}
        \bm{J}_{\bm{p}\bm{p}} & \bm{J}_{\bm{p}\bm{\epsilon}} \\
        \bm{J}_{\bm{\epsilon}\bm{p}} & \bm{J}_{\bm{\epsilon}\bm{\epsilon}}
    \end{bmatrix}, \label{eq:fim_small_structure}
\end{align}
where sub-blocks $\bm{J}_{\bm{\zeta}_i \bm{\zeta}_j}$ are computed from partial derivatives of the mean signal model
\begin{align}
    \bm{\mu}_k = \bm{a}_k(\bm{p}, \bm{\psi}) \tilde{s}_k \label{eq:mu_k}
\end{align}
with respect to position $\bm{p}$, beampattern parameters $\bm{\psi}$, and signal $\tilde{\bm{s}}$. Full derivations are given in Section~S1 of the supplementary material.

\subsection{Position Bound}

Using the partitioned matrix inversion formula \cite{kayFundamentalsStatisticalSignal1993}, the CRLB for position can be extracted from the full FIM by
\begin{align}
    \text{CRLB}_{\bm{p}} &= \frac{\sigma_n^2}{2}\left[\bm{J}_{\bm{p}\bm{p}} - \bm{J}_{\bm{p}\bm{\epsilon}} \bm{J}\inv_{\bm{\epsilon}\bm{\epsilon}} \bm{J}_{\bm{\epsilon}\bm{p}}\right]\inv. \label{eq:crlb_position}
\end{align}

The scalar position error bound is then
\begin{align}
    \sigma_{\bm{p}} &= \sqrt{\tr(\text{CRLB}_{\bm{p}})}. \label{eq:position_bound}
\end{align}

\begin{revisedblock}
\subsection{Geometric Interpretation} \label{subsec:geometry}
The bound depends on receiver geometry through the beampattern gradient. As the emitter moves, the transmit angle to receiver $l$ changes, and with it the gain the receiver sees. Differentiating (\ref{eq:g_def}),
\begin{align}
    \frac{\partial g_l(\bm{p}, \bm{\psi})}{\partial \theta_l^{(t)}} &= -\alpha(\beta)\, g_l(\bm{p}, \bm{\psi}) \sin\!\left(\theta_l^{(t)}(\bm{p})-\phi\right), \label{eq:gain_gradient}
\end{align}
which vanishes at boresight where the lobe is flat and peaks on its steep flank. Near the peak a receiver sees almost no change in received power as the emitter moves, and constrains position through delay and array phase alone. On the flank it registers a large swing in power, supplying RSS information that tightens the bound. The gradient also scales as $\norm{\bm{p}-\bm{u}_l}\inv$, favouring receivers close to the emitter.

The orientation and beamwidth are unknown apriori, so this sensitivity cannot be used to sit a receiver on a known lobe edge. However, it may be used to inform the search formation of receivers. Spreading them across a wide span of aspect angles ensures some fall on a flank however the beam is pointed, and requires no prior. Clustering them near an assumed boresight is fragile: a small error in the assumed direction strands the whole formation on the flat peak. The prescription concerns the shape of the formation, not any individual receiver, and the estimator needs no beam prior. \Cref{subsec:beampattern_sensitivity} quantifies both claims.
\end{revisedblock}

\section{Directional Emitter Localization Algorithm} \label{sec:dpdalg}

In this section we develop a directional emitter-aware DPD cost function for localization.

\subsection{Cost Function}
Following the standard maximum likelihood (ML) approach, we minimize the negative log-likelihood function (\ref{eq:log-likelihood}), which is equivalent to minimizing
\begin{align}
    Q(\bm{p}, \bm{\psi}) &= \sum_{k=0}^{N_s-1}\norm{\tilde{\bm{r}}_k-\bm{a}_k(\bm{p}, \bm{\psi})\tilde{s}_k}^2. \label{eq:inital_cost}
\end{align}
The signal coefficients $\tilde{s}_k$ are nuisance parameters. The value of $\tilde{s}_k$ that minimizes (\ref{eq:inital_cost}) for a given $(\bm{p}, \bm{\psi})$ is the pseudoinverse solution
\begin{align}
    \hat{\tilde{s}}_k &= [\bm{a}_k(\bm{p}, \bm{\psi})]^\dagger \tilde{\bm{r}}_k \\
    &= \frac{[\bm{a}_k(\bm{p}, \bm{\psi})]^\H}{M \norm{\bm{d}(\bm{p}, \bm{\psi})}^2}\, \tilde{\bm{r}}_k, \label{eq:pseudo_def}
\end{align}
where $\bm{d}(\bm{p}, \bm{\psi}) \triangleq [d_1(\bm{p}, \bm{\psi}) \dots{} d_L(\bm{p}, \bm{\psi})]^\T$ collects the directional path attenuations. The simplification follows because the elements of the receiver steering vectors $\bm{a}_l(\bm{p})$ in (\ref{eq:steering_vector}) have unit magnitude, so $[\bm{a}_k]^\H \bm{a}_k = M \norm{\bm{d}}^2$ with all phase terms canceling.

Substituting (\ref{eq:pseudo_def}) into (\ref{eq:inital_cost}) and expanding,
\begin{align}
    Q(\bm{p}, \bm{\psi}) &= \sum_{k=0}^{N_s-1} \norm{\tilde{\bm{r}}_k}^2 - \frac{\sum_{k=0}^{N_s-1} \left|[\bm{a}_k(\bm{p}, \bm{\psi})]^\H \tilde{\bm{r}}_k\right|^2}{M \norm{\bm{d}(\bm{p}, \bm{\psi})}^2}. \label{eq:expanded_cost}
\end{align}
Since $\norm{\tilde{\bm{r}}_k}^2$ is independent of the candidate parameters, minimizing $Q(\bm{p}, \bm{\psi})$ is equivalent to maximizing
\begin{align}
    \tilde{Q}(\bm{p}, \bm{\psi}) &= \sum_{k=0}^{N_s-1} \frac{[\bm{a}_k(\bm{p}, \bm{\psi})]^\H \bm{R}_k\, \bm{a}_k(\bm{p}, \bm{\psi})}{\norm{\bm{d}(\bm{p}, \bm{\psi})}^2}, \label{eq:qtildesol}
\end{align}
where $\bm{R}_k \triangleq \tilde{\bm{r}}_k \tilde{\bm{r}}_k^\H$ is the outer product at frequency bin $k$. This is the matched Bartlett cost function: the numerator measures phase coherence across receivers and frequencies, and the denominator normalizes by the total RSS so that the cost depends on the relative amplitude distribution rather than absolute power.

In the general DPD formulation, unknown per-receiver channel gains are concentrated out via eigenvalue maximization~\cite{weissDirectPositionDetermination2004}, which is appropriate when the cost function depends only on phase information. However, when RSS information is exploited, unknown channel gains are indistinguishable from the directional attenuation $d_l(\bm{p}, \bm{\psi})$, and the eigenvalue approach systematically flattens the perceived beampattern~\cite{salmanJointEstimationRSSBased2012}. Recent work incorporates path loss models that constrain channel gain ratios as a function of position to regularize this ambiguity~\cite{youJointDirectEstimation2023, zhaoContextualDirectPosition2025,youDirectPositionDetermination2026}.

The emitter position and beampattern parameters are then estimated by
\begin{align}
    (\hat{\bm{p}}, \hat{\bm{\psi}}) = \argmax_{\bm{p}, \bm{\psi}} \tilde{Q}(\bm{p}, \bm{\psi}). \label{eq:psolve}
\end{align}
Direct evaluation of (\ref{eq:psolve}) requires a four-dimensional grid search over $(\bm{p}, \bm{\psi})$, \revised{whose cost scales with the product of the position and beampattern grid sizes (\cref{subsec:complexity})}. We present two methods to address this: a block coordinate ascent algorithm that iteratively alternates between position and beampattern subproblems, and a profile likelihood method that analytically concentrates the beampattern parameters out of the cost function at each candidate position.

\subsection{Block Coordinate Ascent (BCA) Algorithm} \label{subsec:bca}
A natural simplification is block coordinate ascent: alternately maximize $\tilde{Q}$ over position $\bm{p}$ with $\bm{\psi}$ fixed, then over beampattern $\bm{\psi}$ with $\bm{p}$ fixed, each step being a two-dimensional search. Since each block maximization is non-decreasing, the sequence converges to a stationary point of $\tilde{Q}(\bm{p}, \bm{\psi})$ \cite{tsengConvergenceBlockCoordinate2001}, \revised{which is in general only a local maximum}. Algorithm~\ref{alg:estimation} details this procedure. The initial position estimate is obtained by setting $d_l = 1$ for all $l$, removing all RSS information and reducing the cost function to phase-only localization. \revised{The convergence behaviour and the resulting low-SNR performance are examined in \cref{subsec:localisation_performance,subsec:convergence}.}

\begin{algorithm}[!t]
    \caption{Block Coordinate Ascent for Joint Position and Beampattern Estimation}
    \label{alg:estimation}
    \KwIn{Received signals $\{\bm{r}_{l}\}_{l=1}^{L}$, receiver positions $\{\bm{u}_{l}\}_{l=1}^{L}$, convergence threshold $\varepsilon$}
    \KwOut{Estimated position $\hat{\bm{p}}$ and beampattern $\hat{\bm{\psi}}$}
    $\hat{\bm{p}} \gets \argmax_{\bm{p}} \tilde{Q}(\bm{p}, \bm{\psi})\big|_{d_l=1}$\;
    $\Delta \gets \infty$\;
    \While{$\Delta > \varepsilon$}{
        $\hat{\bm{p}}_{\text{prev}} \gets \hat{\bm{p}}$\;
        $\hat{\bm{\psi}} \gets \argmax_{\bm{\psi}} \tilde{Q}(\hat{\bm{p}}, \bm{\psi})$\;
        $\hat{\bm{p}} \gets \argmax_{\bm{p}} \tilde{Q}(\bm{p}, \hat{\bm{\psi}})$\;
        $\Delta \gets \norm{\hat{\bm{p}} - \hat{\bm{p}}_{\text{prev}}}$\;
    }
    \Return{$\hat{\bm{p}}, \hat{\bm{\psi}}$}\;
\end{algorithm}

\subsection{Profile Likelihood Algorithm} \label{subsec:profile_likelihood}
The block coordinate ascent approach has two limitations: the beampattern subproblem still requires a grid search over $\bm{\psi}$ at each iteration, and the solution depends on initialization quality, which degrades at low SNR. Both are addressed by analytically concentrating the beampattern parameters out of the cost function at each candidate position, producing a position-only search surface called the profile likelihood
\begin{align}
    \tilde{Q}_{\text{PL}}(\bm{p}) &= \max_{\bm{\psi}} \tilde{Q}(\bm{p}, \bm{\psi}). \label{eq:profile_def}
\end{align}
This reduces the four-dimensional joint optimization to a two-dimensional grid search over position, where the inner maximization over $\bm{\psi}$ is \revised{carried out independently at each grid point by the low-dimensional procedure developed below}.

At a fixed candidate position $\bm{p}$, the transmit angles $\theta_l^{(t)}(\bm{p})$ and isotropic path attenuations $\rho_l(\bm{p})$ from (\ref{eq:rho_def}) are known for all $l$. Define
\begin{align}
    \bm{v}_{k,l}(\bm{p}) &\triangleq \rho_l(\bm{p})\, e^{-j\omega_k \tau_l(\bm{p})}\, \bm{a}_l(\bm{p}), \label{eq:v_def} \\
    \bm{V}_k(\bm{p}) &\triangleq \revised{\operatorname{blkdiag}\!\big(\bm{v}_{k,1}(\bm{p}), \dots{}, \bm{v}_{k,L}(\bm{p})\big)}, \label{eq:V_def} \\
    \bm{g}(\bm{p}, \bm{\psi}) &\triangleq [g_1(\bm{p}, \bm{\psi}) \dots{} g_L(\bm{p}, \bm{\psi})]^\T, \\
    \bm{\rho}(\bm{p}) &\triangleq [\rho_1(\bm{p}) \dots{} \rho_L(\bm{p})]^\T,
\end{align}
\revised{where $\bm{v}_{k,l}(\bm{p})$ is the $M \times 1$ phase-steered vector for the $l$th receiver at frequency $k$, carrying every position-dependent factor except the beampattern gain, and $\bm{V}_k(\bm{p})$ is the resulting $ML \times L$ block-diagonal matrix. Since $d_l = \rho_l\, g_l$ by (\ref{eq:d_def}), the stacked steering vector (\ref{eq:a_def}) factors as $\bm{a}_k(\bm{p}, \bm{\psi}) = \bm{V}_k(\bm{p})\, \bm{g}(\bm{p}, \bm{\psi})$, and substituting into (\ref{eq:qtildesol}) gives a generalized Rayleigh quotient}
\begin{align}
    \tilde{Q}(\bm{p}, \bm{\psi}) &= \frac{\bm{g}(\bm{p}, \bm{\psi})^\T \tilde{\bm{C}}(\bm{p})\, \bm{g}(\bm{p}, \bm{\psi})}{\bm{g}(\bm{p}, \bm{\psi})^\T \bm{W}(\bm{p})\, \bm{g}(\bm{p}, \bm{\psi})}, \label{eq:rayleigh_quotient}
\end{align}
where
\begin{align}
    \bm{C}(\bm{p}) &= \sum_{k=0}^{N_s-1} \bm{V}_k^\H(\bm{p})\, \bm{R}_k\, \bm{V}_k(\bm{p}), \label{eq:C_def} \\
    \tilde{\bm{C}}(\bm{p}) &= \operatorname{Re}\{\bm{C}(\bm{p})\}, \\
    \bm{W}(\bm{p}) &= \operatorname{diag}(\bm{\rho}(\bm{p}) \odot \bm{\rho}(\bm{p})).
\end{align}
Since $\bm{g}(\bm{p}, \bm{\psi})$ is real-valued, only the real part of $\bm{C}(\bm{p})$ contributes to the quadratic form.

The unconstrained maximum of (\ref{eq:rayleigh_quotient}) over all $\bm{g}$ (not restricted to the parametric beampattern family) is the largest generalized eigenvalue of the pencil $(\tilde{\bm{C}}, \bm{W})$:
\begin{align}
    \tilde{\bm{C}}(\bm{p})\, \bar{\bm{g}} &= \lambda_{\max}(\bm{p})\, \bm{W}(\bm{p})\, \bar{\bm{g}}. \label{eq:gen_eig}
\end{align}
Since $\bm{W}$ is diagonal and positive definite, this reduces to a standard $L \times L$ real symmetric eigenvalue problem, inexpensive for typical array sizes. The eigenvector $\bar{\bm{g}}$ is the amplitude pattern the data favors at $\bm{p}$ with no beampattern constraint, and $\lambda_{\max}(\bm{p})$ upper-bounds $\tilde{Q}(\bm{p}, \bm{\psi})$ for any beam parameters.

The eigenvector $\bar{\bm{g}}$ will not in general correspond to any $\bm{\psi}$ in the parametric model. Since the unconstrained amplitude pattern at receiver $l$ is $\rho_l(\bm{p})\, \bar{g}_l$ and the model predicts $\rho_l(\bm{p})\, g_l(\bm{p}, \bm{\psi})$, factoring out $\rho_l$ yields the weighted least-squares fit
\begin{align}
    \hat{\bm{\psi}} &= \argmin_{\bm{\psi}} \sum_{l=1}^{L} \rho_l^2(\bm{p}) \left(\bar{g}_l - g_l(\bm{p}, \bm{\psi})\right)^2, \label{eq:bp_fit}
\end{align}
where the weights $\rho_l^2(\bm{p})$ arise naturally from the amplitude decomposition. \revised{This two-dimensional problem has no closed-form solution in $\bm{\psi}$, but a closed-form \emph{approximate} solution, used as a seed for the Newton refinement below, does. Taking logs of the von Mises gain,
\begin{align}
    \ln g_l(\bm{p}, \bm{\psi}) = \alpha(\beta)\!\left[\cos(\theta_l^{(t)}(\bm{p}) - \phi) - 1\right] = \bm{a}^\T \bm{x}_l, \label{eq:log_linear}
\end{align}
which is linear in the coefficient and regressor vectors
\begin{align}
    \bm{a} &= [\alpha\cos\phi,\ \alpha\sin\phi,\ -\alpha]^\T, \label{eq:seed_coeff} \\
    \bm{x}_l &= [\cos\theta_l^{(t)}(\bm{p}),\ \sin\theta_l^{(t)}(\bm{p}),\ 1]^\T. \label{eq:seed_regressor}
\end{align}
Fitting $\ln\bar{g}_l$ to $\bm{x}_l$ by weighted linear least squares, with log-domain weights $w_l = \rho_l^2(\bm{p})\,\bar{g}_l^2$, gives the coefficients in closed form as the solution of a $3\times3$ system,
\begin{align}
    \hat{\bm{a}} = \Bigl(\sum_{l} w_l\, \bm{x}_l \bm{x}_l^\T\Bigr)\inv \sum_{l} w_l\, \bm{x}_l \ln\bar{g}_l, \label{eq:seed_normal}
\end{align}
computed in $\mathcal{O}(L)$ with no grid. The seed is recovered as
\begin{align}
    \hat{\phi}_0 = \arctantwo(\hat{a}_2, \hat{a}_1), \qquad \alpha(\hat{\beta}_0) = \sqrt{\hat{a}_1^2 + \hat{a}_2^2}. \label{eq:seed_recover}
\end{align}
Dropping the exact constraint $\hat{a}_3 = -\sqrt{\hat{a}_1^2 + \hat{a}_2^2}$ is what makes the fit linear; the Newton step restores it. When the receivers subtend too narrow a range of transmit angles for this fit to be well conditioned, we fall back to enumerating the $(\phi, \beta)$ candidate set used by BCA, which maximizes the scale-concentrated objective
\begin{align}
    \hat{\bm{\psi}}_0 &= \argmax_{\bm{\psi}}
    \frac{\left(\sum_l \rho_l^2\, \bar{g}_l\, g_l(\bm{p}, \bm{\psi})\right)^2}
         {\sum_l \rho_l^2\, g_l^2(\bm{p}, \bm{\psi})}, \label{eq:bp_fit_sep}
\end{align}
at $L$ operations per candidate. The two seeds reach stationary points of equivalent localization accuracy (Section~S4), so we use the closed-form seed by default for its lower cost (\cref{subsec:runtime}).}

The profile likelihood at position $\bm{p}$ is then $\tilde{Q}_{\text{PL}}(\bm{p}) = \tilde{Q}(\bm{p},\, \hat{\bm{\psi}}(\bm{p}))$, evaluated via (\ref{eq:rayleigh_quotient}) with $\bm{g} = \bm{g}(\bm{p}, \hat{\bm{\psi}})$.

The WLS fit (\ref{eq:bp_fit}) is an approximate inner solve: it minimises a weighted squared error on the amplitude pattern rather than directly maximising the constrained Rayleigh quotient. \revised{This warm start is refined towards the constrained maximiser by Newton iteration on $\tilde{Q}(\bm{p}, \bm{\psi})$, using the analytic gradient and Hessian derived in Section~S3 of the supplementary material. The refinement is safeguarded: when the Hessian is not negative definite the Newton step falls back to the normalized gradient, with step length set by Armijo backtracking. Every accepted step then increases $\tilde{Q}(\bm{p}, \cdot)$, bounded above by $\lambda_{\max}(\bm{p})$ via (\ref{eq:gen_eig}), so the iteration is monotone and cannot diverge, reaching a local maximum near the warm start (\cref{subsec:convergence}).} We adopt the Newton-refined inner solve as the default throughout the remainder of the paper.

\begin{algorithm}[!t]
    \caption{Profile Likelihood for Directional Emitter Localization}
    \label{alg:profile_likelihood}
    \KwIn{Received signals $\{\bm{r}_{l}\}_{l=1}^{L}$, receiver positions $\{\bm{u}_{l}\}_{l=1}^{L}$, position grid $\mathcal{G}_{\bm{p}}$}
    \KwOut{Estimated position $\hat{\bm{p}}$ and beampattern $\hat{\bm{\psi}}$}
    \For{each $\bm{p} \in \mathcal{G}_{\bm{p}}$}{
        Compute $\rho_l(\bm{p})$ and $\theta_l^{(t)}(\bm{p})$ for all $l$\;
        Form $\tilde{\bm{C}}(\bm{p})$ and $\bm{W}(\bm{p})$ via (\ref{eq:C_def})\;
        \tcp{Unconstrained optimum}
        Solve $\tilde{\bm{C}}(\bm{p})\, \bar{\bm{g}} = \lambda_{\max}\, \bm{W}(\bm{p})\, \bar{\bm{g}}$\;
        \tcp{Seed by closed-form log-domain fit\\ (enumerate if ill-conditioned)}
        $\hat{\bm{\psi}}_0 \gets$ log-linear WLS on $\ln\bar{\bm{g}}$\;
        \tcp{Optimize constrained\\ Rayleigh quotient}
        $\hat{\bm{\psi}} \gets$ Newton$\left(\tilde{Q}(\bm{p}, \cdot);\, \hat{\bm{\psi}}_0\right)$\;
        $\tilde{Q}_{\text{PL}}(\bm{p}) \gets \tilde{Q}(\bm{p}, \hat{\bm{\psi}})$ via (\ref{eq:rayleigh_quotient})\;
    }
    $\hat{\bm{p}} \gets \argmax_{\bm{p}} \tilde{Q}_{\text{PL}}(\bm{p})$\;
    Set $\hat{\bm{\psi}}$ to the $\hat{\bm{\psi}}$ computed at $\hat{\bm{p}}$\;
    \Return{$\hat{\bm{p}}, \hat{\bm{\psi}}$}\;
\end{algorithm}

Algorithm~\ref{alg:profile_likelihood} summarizes the procedure. Unlike the alternation in BCA (Algorithm~\ref{alg:estimation}), it evaluates each position grid point independently with a single eigendecomposition and a low-dimensional inner refinement, making it naturally parallelizable.

\subsection{Computational Complexity} \label{subsec:complexity}
\begin{revisedblock}
All three methods share the same per-position precomputation. At a candidate $\bm{p}$, beamforming the $L$ receivers over $N_s$ frequency bins costs $\mathcal{O}(L M N_s)$ and forming $\tilde{\bm{C}}(\bm{p})$ in (\ref{eq:C_def}) costs $\mathcal{O}(L^2 N_s)$, giving
\begin{align}
    C_{\text{pre}} &= \mathcal{O}\!\left(L N_s (M + L)\right). \label{eq:cost_pre}
\end{align}
Once $\tilde{\bm{C}}$ is available, evaluating the cost for a given $\bm{\psi}$ is only the quadratic form $\mathcal{O}(L^2)$, so the methods differ in how often, and at which positions, they perform the beampattern search.

An exhaustive search evaluates every $\bm{\psi}$ candidate at every position,
\begin{align}
    C_{\text{ex}} &= \mathcal{O}\!\left(|\mathcal{G}_{\bm{p}}|\left[C_{\text{pre}} + |\mathcal{G}_{\bm{\psi}}| L^2\right]\right). \label{eq:cost_exhaustive}
\end{align}
The profile likelihood also visits every position but does not enumerate $\bm{\psi}$: it forms the closed-form log-domain seed in $\mathcal{O}(L)$, takes one $\mathcal{O}(L^3)$ eigendecomposition, and runs $N_{\text{n}}$ Newton steps,
\begin{align}
    C_{\text{PL}} &= \mathcal{O}\!\left(|\mathcal{G}_{\bm{p}}|\left[C_{\text{pre}} + L^3 + N_{\text{n}}(1 + N_{\text{ls}}) L^2\right]\right), \label{eq:cost_pl}
\end{align}
where $N_{\text{ls}}$ is the average number of backtracking evaluations per Newton step; the enumeration fallback, when triggered, adds an $\mathcal{O}(|\mathcal{G}_{\bm{\psi}}| L)$ seed term. BCA differs structurally: it searches $\bm{\psi}$ at a \emph{single} position per alternation rather than at every grid point, so the beampattern grid enters only once per iteration,
\begin{align}
    C_{\text{BCA}} &= \mathcal{O}\!\left(N_{\text{it}}\left[|\mathcal{G}_{\bm{p}}| C_{\text{pre}} + |\mathcal{G}_{\bm{\psi}}| L^2\right]\right). \label{eq:cost_bca}
\end{align}

The exhaustive search carries the beampattern grid at every position, costing $|\mathcal{G}_{\bm{p}}||\mathcal{G}_{\bm{\psi}}|$. The profile likelihood and BCA both avoid this product: the profile likelihood replaces the grid with the closed-form seed and continuous Newton refinement, scaling with $|\mathcal{G}_{\bm{p}}|$ alone, and BCA searches $\bm{\psi}$ at one position per iteration, scaling with $N_{\text{it}}(|\mathcal{G}_{\bm{p}}| + |\mathcal{G}_{\bm{\psi}}|)$. BCA is the cheapest of the three at the small $N_{\text{it}}$ seen in practice.

The $|\mathcal{G}_{\bm{\psi}}|$ factor is what separates the exhaustive search from the profile likelihood as accuracy demands grow. For the exhaustive search the grid \emph{is} the estimate: its $\bm{\psi}$-precision is the grid spacing, so improving it forces $|\mathcal{G}_{\bm{\psi}}|$ to grow and, by (\ref{eq:cost_exhaustive}), the cost with it. The profile likelihood resolves $\bm{\psi}$ continuously through the Newton step, so its accuracy is decoupled from any grid. At a coarse shared grid the profile likelihood is already the cheaper of the two, and the gap widens without bound as the exhaustive grid is refined. This is the sense in which the four-dimensional joint search is avoided; measured runtimes are reported in \cref{subsec:runtime}.
\end{revisedblock}

\section{Simulation Results} \label{sec:sim}

\subsection{Experimental Design}
We evaluate the proposed algorithms through Monte Carlo simulation of two representative directional-emitter scenarios, communications and radar, using the root mean square (RMS) distance error between the estimated and true emitter position as the localization metric. The baseline is the AOA--TDOA DPD method of Weiss \cite{weissDirectPositionDetermination2004}, which does not account for directional emission. Common simulation parameters are summarized in Table~\ref{tab:sim_params}.

\begin{table}[!t]
    \centering
    \caption{Simulation parameters common to both scenarios.}
    \label{tab:sim_params}
    \begin{tabular}{ll}
        \hline
        Parameter & Value \\
        \hline
        Receiver array elements ($M$) & 4 (ULA) \\
        Samples per snapshot ($N_s$) & \revised{64} \\
        Sampling frequency & \revised{200~kHz} \\
        SNR range & $-15$ to $0$~dB, step $1$~dB \\
        Orientation grid & \revised{$5\degree$ resolution, $[-180\degree, 180\degree)$} \\
        Beamwidth grid & \revised{$2.5\degree$ resolution, $[10\degree, 90\degree]$} \\
        Monte Carlo trials per SNR & 250 \\
        \hline
    \end{tabular}
\end{table}

Per-scenario geometry is summarized in Table~\ref{tab:scenarios} and visualized in Fig.~\ref{fig:experimental_setup}. The best-case SNR (at the receiver with the strongest signal) is the value varied; because each receiver experiences different RSS due to beampattern and path attenuation, the realized per-receiver SNR differs across the formation.

\begin{table}[!t]
    \centering
    \caption{Per-scenario emitter and receiver geometry.}
    \label{tab:scenarios}
    \small
    \begin{tabular}{@{}lll@{}}
        \hline
        Parameter & Communications & Radar \\
        \hline
        Emitter position $\bm{p}$ (m) & $[600, 600]$ & $[-4000, 4000]$ \\
        Orientation $\phi$ & $-10\degree$ & $-45\degree$ \\
        Beamwidth $\beta$ & $30\degree$ & $30\degree$ \\
        Receivers ($L$) & 4 corners $\pm 2500$~m & 4 spread (Fig.~\ref{fig:directive_radar_setup}) \\
        \revised{Position grid (multi-stage)} & \revised{$50 \to 1$~m} & \revised{$20 \to 1$~m} \\
        \hline
    \end{tabular}
\end{table}

\input{img/figures/experimental_setup}

The communications scenario (Fig.~\ref{fig:directive_comms_setup}) represents a typical geometry where the emitter sits inside a four-corner receiver formation and only a subset of receivers are illuminated by the main lobe. The radar scenario (Fig.~\ref{fig:directive_radar_setup}) represents a forward-deployed sensor formation tasked with localizing a directional radar from positions that span the beampattern: \revised{$R_1$ and $R_2$ lie within a few degrees of boresight and therefore sit near peak gain, while $R_3$ and $R_4$ lie roughly $25\degree$ and $20\degree$ off boresight, beyond the $15\degree$ half-power edge and into the low-gain tail. Each receiver thus samples a distinct point of the gain profile, giving a strong RSS contrast across the formation.}

\subsection{Localization Performance} \label{subsec:localisation_performance}
We compare the AOA--TDOA baseline \cite{weissDirectPositionDetermination2004}, BCA (Algorithm~\ref{alg:estimation}), the profile likelihood (Algorithm~\ref{alg:profile_likelihood}), and an oracle-$\beta$ baseline that fixes the true beamwidth and alternates only the orientation $\phi$. The oracle-$\beta$ baseline is a best-case version of the prior-art approach of \cite{youJointDirectEstimation2023}, which assumes a catalog beamwidth prior: supplying the exact $\beta$ removes any prior-mismatch penalty from the comparison.

\begin{figure*}[!t]
    \centering
    \setlength{\tempwidth}{0.28\textwidth}
    \settoheight{\tempheight}{\includegraphics[width=0.28\textwidth]{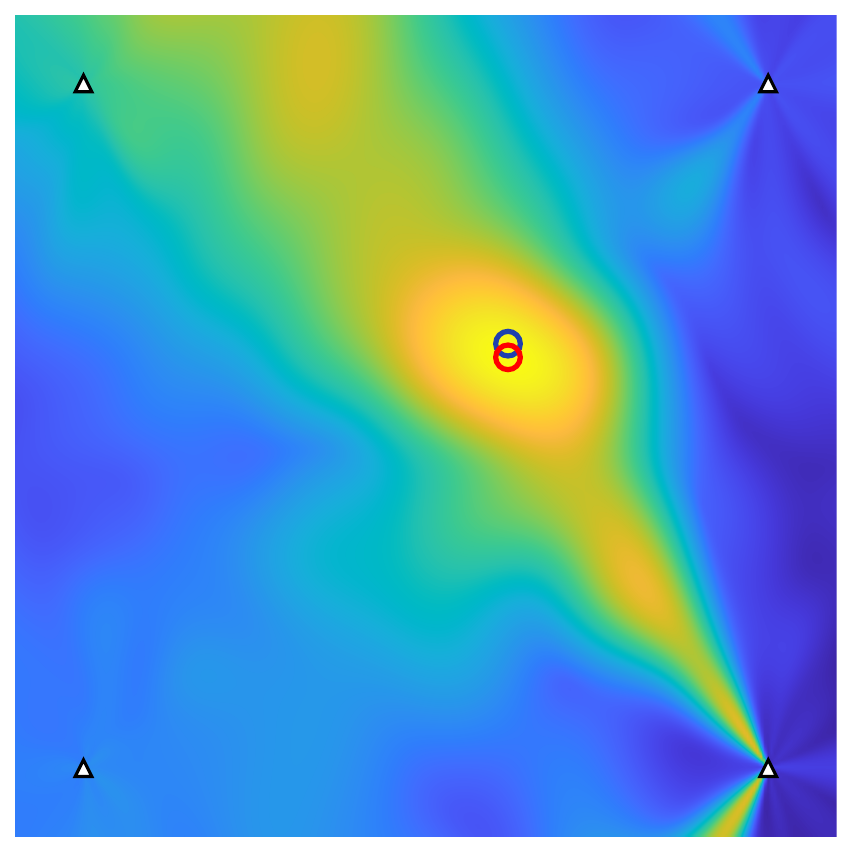}}%
    \columnname{AOA--TDOA}\hfill%
    \columnname{BCA (Proposed)}\hfill%
    \columnname{PL (Proposed)}\\
    \vspace{-0.8em}
    \rowname{Comms.}
    \subfloat[]{\label{fig:qualitative-comp-a}%
    \includegraphics[width=0.28\textwidth, height=\tempheight]{img/figures/heatmaps/comms_aoa_tdoa.pdf}}%
    \subfloat[]{\label{fig:qualitative-comp-b}%
    \includegraphics[width=0.28\textwidth, height=\tempheight]{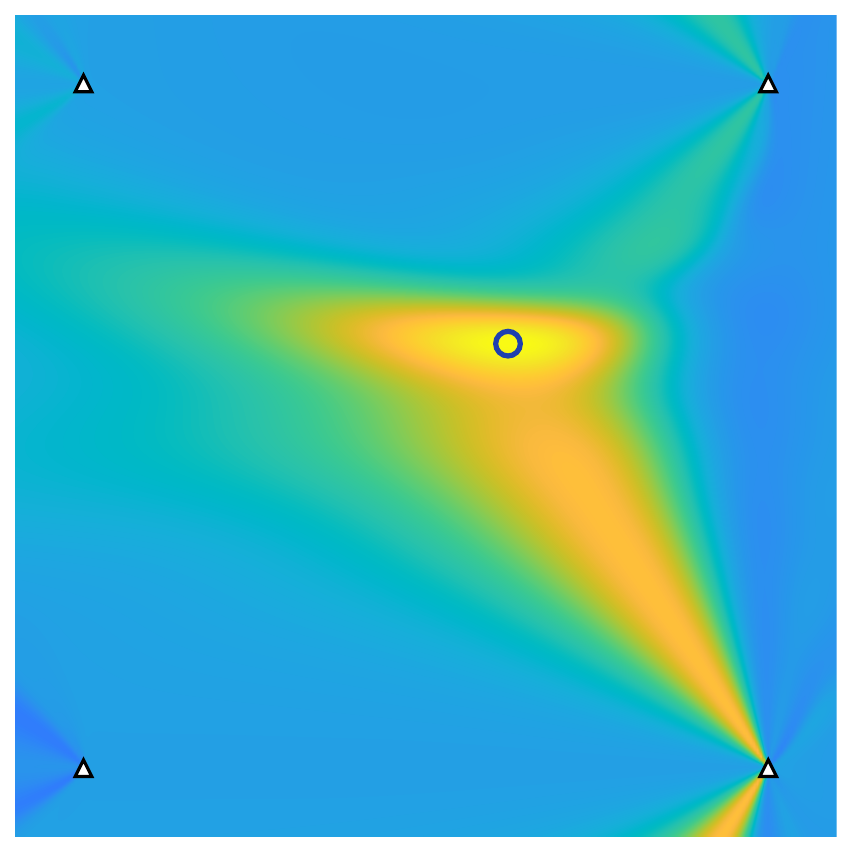}}%
    \subfloat[]{\label{fig:qualitative-comp-c}%
    \includegraphics[width=0.28\textwidth, height=\tempheight]{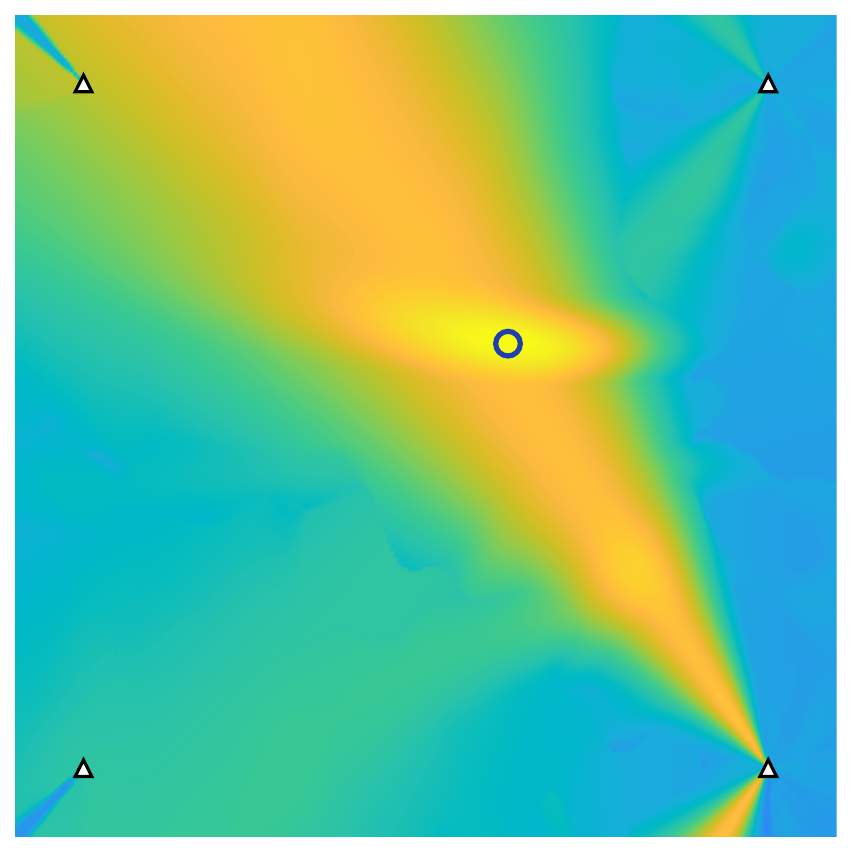}}%
    \\[-1em]
    \rowname{Radar}
    \subfloat[]{\label{fig:qualitative-comp-d}%
    \includegraphics[width=0.28\textwidth, height=\tempheight]{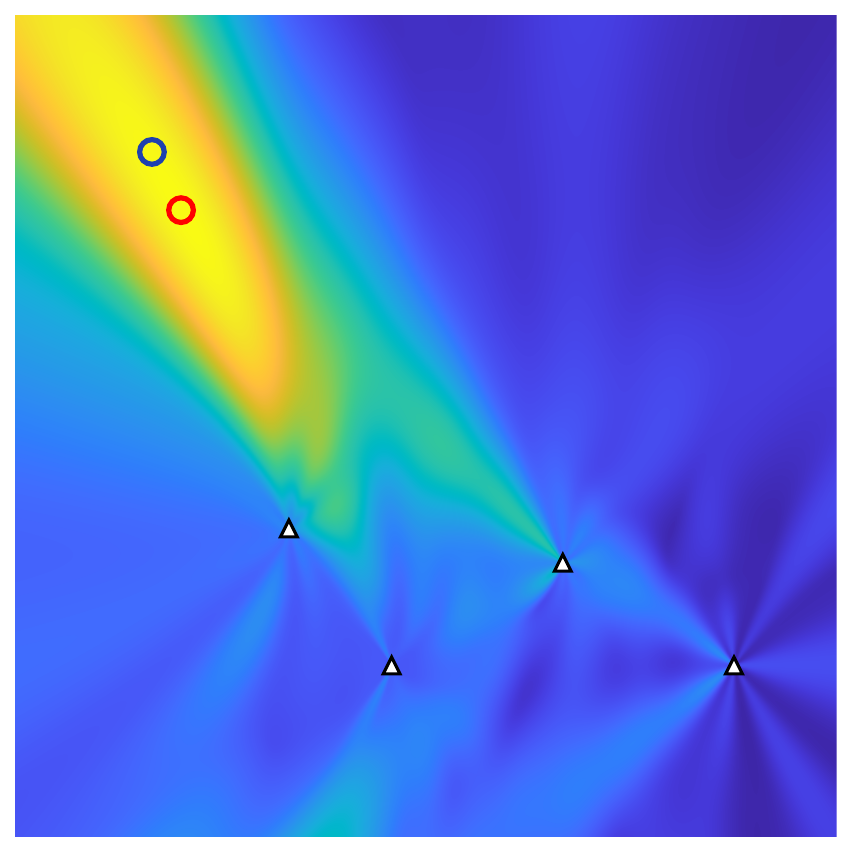}}%
    \subfloat[]{\label{fig:qualitative-comp-e}%
    \includegraphics[width=0.28\textwidth, height=\tempheight]{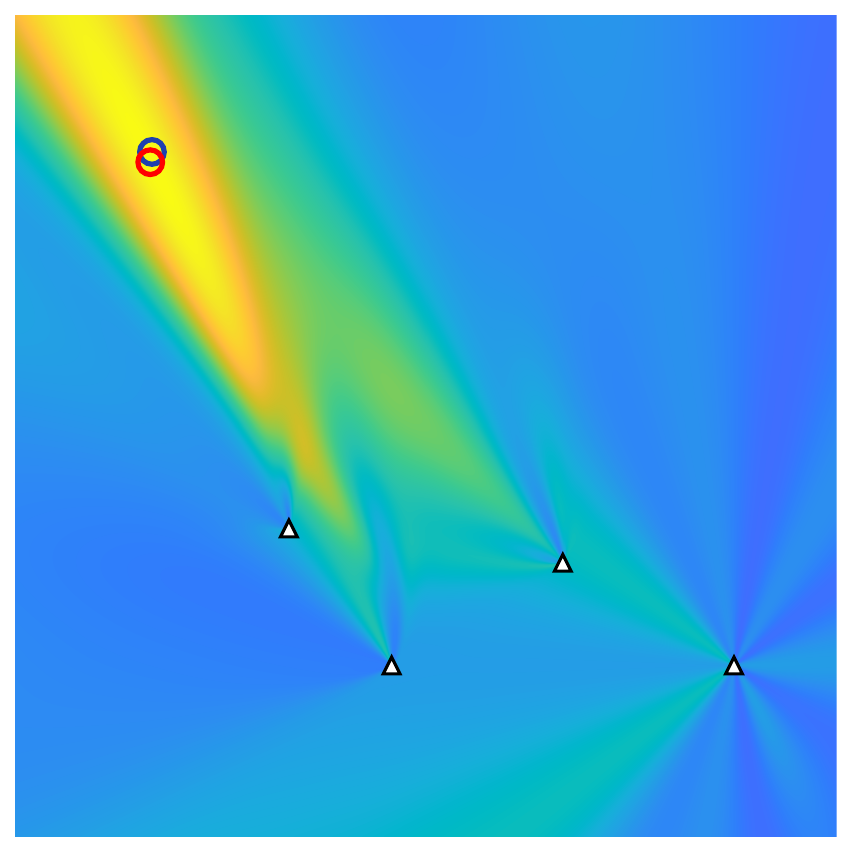}}%
    \subfloat[]{\label{fig:qualitative-comp-f}%
    \includegraphics[width=0.28\textwidth, height=\tempheight]{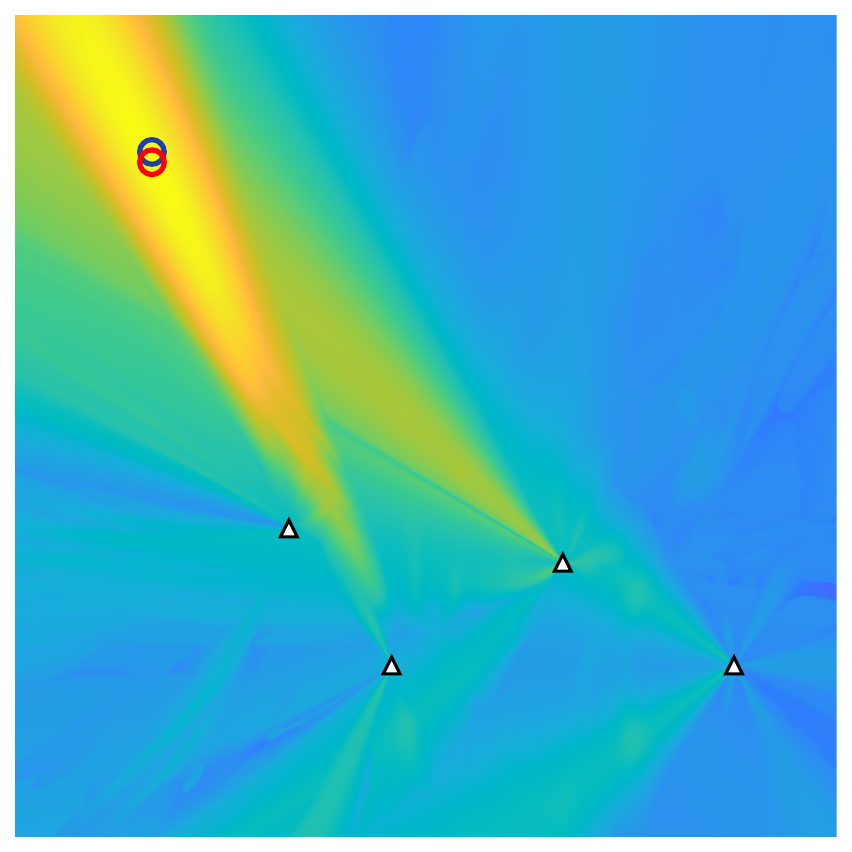}}%
    \caption{Cost-function heatmaps for one representative realization at SNR $=-5$~dB ($N_s=64$, $M=4$). The blue circle marks the true emitter, the red circle each method's cost-surface peak, and triangles the receivers. \revised{In the communications geometry every method localizes the emitter at this SNR; in the radar geometry AOA--TDOA is drawn along the delay-ambiguity ridge away from the emitter, while the beampattern-aware methods exploit the RSS modulation to localize it.}}
    \label{fig:qualitative_heatmaps_grid}
\end{figure*}

Fig.~\ref{fig:qualitative_heatmaps_grid} presents cost-function heatmaps for one representative realization at $-5$~dB. In the communications experiment the four-corner geometry is well conditioned at this SNR, and all three methods place their peak at the emitter: AOA--TDOA (Fig.~\ref{fig:qualitative-comp-a}) over a broad surface along the beam direction, BCA (Fig.~\ref{fig:qualitative-comp-b}) as a tight peak, and the profile likelihood (Fig.~\ref{fig:qualitative-comp-c}) over a broader ridge at the same location.

The radar experiment is harder, and the methods separate. AOA--TDOA (Fig.~\ref{fig:qualitative-comp-d}) gives a broad fan-shaped ridge along the bearing to the emitter and its peak is drawn off the true position, gaining nothing from where the receivers sit within the beampattern, whereas BCA (Fig.~\ref{fig:qualitative-comp-e}) and the profile likelihood (Fig.~\ref{fig:qualitative-comp-f}) exploit the RSS modulation to localize the peak near the true position, tight and broad-ridged respectively.

\begin{figure*}[!t]
    \centering
    \subfloat[Communications, von Mises emitter.]{\label{fig:directive_comms_err_vs_snr}
    \includegraphics[width=7cm]{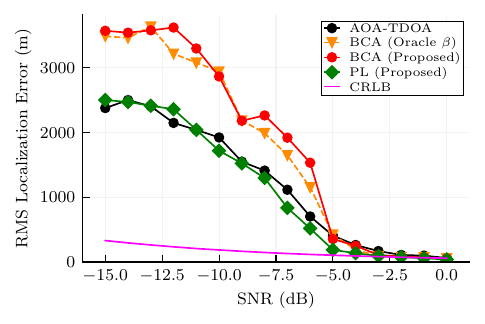}}
    \hfil
    \subfloat[Radar, von Mises emitter.]{\label{fig:directive_radar_err_vs_snr}
    \includegraphics[width=7cm]{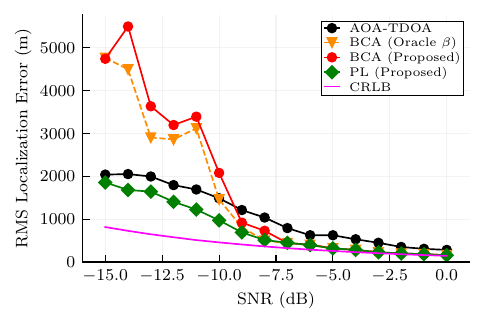}}
    \\
    \subfloat[Communications, ULA emitter (mismatch).]{\label{fig:directive_comms_err_vs_snr_ula}
    \includegraphics[width=7cm]{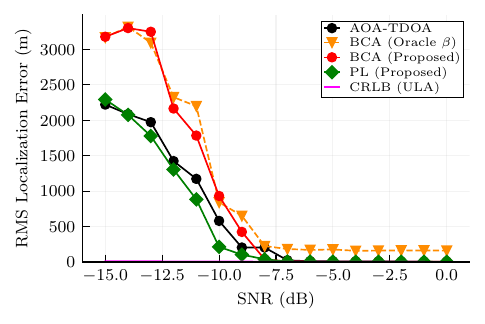}}
    \hfil
    \subfloat[Radar, ULA emitter (mismatch).]{\label{fig:directive_radar_err_vs_snr_ula}
    \includegraphics[width=7cm]{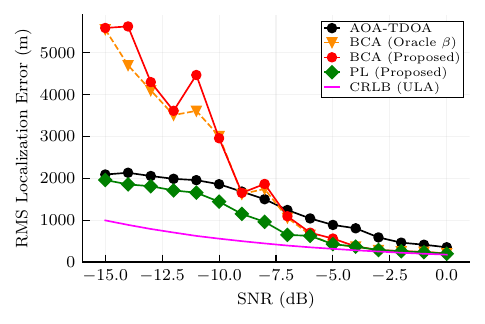}}
    \caption{RMS localization error vs. SNR, aggregated over 250 Monte Carlo trials. Top row: matched model, with simulator and estimator both using the von Mises beampattern. \revised{The profile likelihood approaches the CRLB across the range and separates most from the baselines in the low-SNR threshold region, while the BCA variants degrade most at low SNR.} Bottom row: model mismatch, with the simulator radiating from a $4$-element $\lambda/2$-spaced ULA while the estimator fits the von Mises model; CRLB-ULA is the bound under the true ULA model (Section~S2 of the supplementary material). \revised{The profile likelihood approaches CRLB-ULA while the oracle-$\beta$ baseline plateaus, its beamwidth fixed at the nominal $30\degree$ rather than the ULA-equivalent $\approx 26\degree$.}}
    \label{fig:err_vs_snr}
\end{figure*}

Fig.~\ref{fig:err_vs_snr} presents RMS error from $-15$ to $0$~dB. In the communications experiment (Fig.~\ref{fig:directive_comms_err_vs_snr}), the profile likelihood tracks the AOA--TDOA baseline at the lowest SNRs, separates clearly in the $-7$ to $-5$~dB threshold regime with up to $50\%$ lower RMS error, and joins the CRLB above $-2$~dB. Both BCA variants do substantially worse at low SNR, where the phase-only initialization degrades and the alternation settles in poor local optima; the profile likelihood avoids this through its non-iterative outer structure. Oracle-$\beta$ BCA tracks standard BCA closely, so knowing the true beamwidth helps only marginally and the dominant difficulty is the orientation search, not the beamwidth.

The radar scenario (Fig.~\ref{fig:directive_radar_err_vs_snr}) shows a sharper separation. The profile likelihood sits within a few hundred metres of the CRLB across the full range, while the AOA--TDOA baseline stays about a factor of two above the bound at every SNR. The BCA variants are again worst at low SNR, with errors of $4$--$5$~km below $-10$~dB before catching up to the profile likelihood near $-8$~dB. \revised{The BCA curves are visibly non-monotonic in this region; this is finite-sample noise rather than a real effect: a $4000$-trial run is strictly monotonic, whereas a $250$-trial estimate produces a spurious low-SNR up-step about half the time, since the heavy-tailed BCA error at low SNR makes its RMSE converge slowly.} The deliberate placement of two receivers near boresight and \revised{two beyond the half-power edge} makes the RSS pattern strongly informative for both orientation and beamwidth, so small position perturbations produce measurable RSS asymmetries that pin down the emitter location.

\begin{revisedblock}
\subsection{Beampattern Parameter Estimation} \label{subsec:beam_estimation}
The position estimates are produced jointly with the beampattern parameters, which we now evaluate directly. Fig.~\ref{fig:beam_mae} reports the mean absolute error (MAE) of the orientation $\phi$ and beamwidth $\beta$ against SNR for BCA and the profile likelihood; we use MAE rather than RMSE because the beamwidth error is heavy-tailed at low SNR, where a weakly identified $\beta$ occasionally produces a gross error that a squared metric would let dominate.

The orientation is estimated well by both methods and converges to a few degrees at high SNR, $5\degree$ in the communications scenario and under $3\degree$ in the radar scenario, with the two methods within about a degree of each other throughout. The beamwidth is the harder parameter, and its identifiability depends strongly on geometry, consistent with the sensitivity analysis of \cref{subsec:beampattern_sensitivity}. In the radar scenario, whose spread formation places receivers across the gain profile, both methods recover $\beta$ to roughly $6\degree$ at $0$~dB and track each other across the range. In the communications scenario, where only a subset of the four-corner formation is illuminated, $\beta$ is poorly constrained: the error remains near $16\degree$ even at $0$~dB, and below $-11$~dB the profile-likelihood estimate saturates near $52\degree$ as the unconstrained amplitude pattern ceases to carry usable beamwidth information.

The two methods trade off differently on the beamwidth. BCA enforces the von Mises family on a bounded grid at every step, capping its error and giving a lower-variance estimate at low SNR, most visibly in the communications scenario where it stays near $26\degree$ while the profile likelihood saturates. The profile likelihood relaxes this parametric constraint in its inner solve, which gives it the position advantage of \cref{subsec:localisation_performance} but a noisier beamwidth estimate where the geometry constrains $\beta$ weakly. Where $\beta$ is well constrained, as in the radar scenario, the two are essentially equal on beamwidth and the profile likelihood keeps its position advantage.

\begin{figure*}[!t]
    \centering
    \setlength{\tempwidth}{0.40\textwidth}
    \settoheight{\tempheight}{\includegraphics[width=0.40\textwidth]{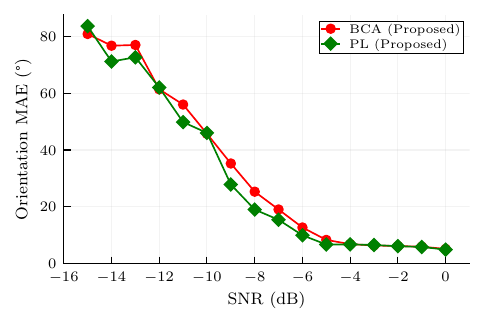}}%
    \columnname{Orientation}\hfill%
    \columnname{Beamwidth}\\
    \vspace{-0.8em}
    \rowname{Comms.}
    \subfloat[Orientation MAE, communications.]{\label{fig:comms_phi_mae}%
    \includegraphics[width=0.40\textwidth, height=\tempheight]{img/figures/comms_phi_rmse.pdf}}%
    \subfloat[Beamwidth MAE, communications.]{\label{fig:comms_beta_mae}%
    \includegraphics[width=0.40\textwidth, height=\tempheight]{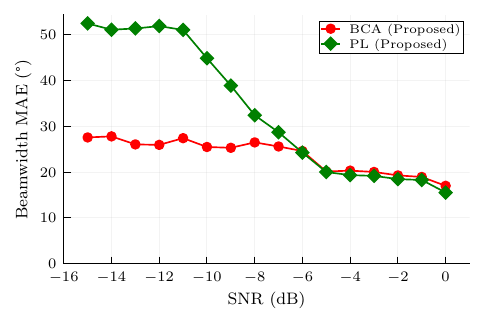}}%
    \\
    \rowname{Radar}
    \subfloat[Orientation MAE, radar.]{\label{fig:radar_phi_mae}%
    \includegraphics[width=0.40\textwidth, height=\tempheight]{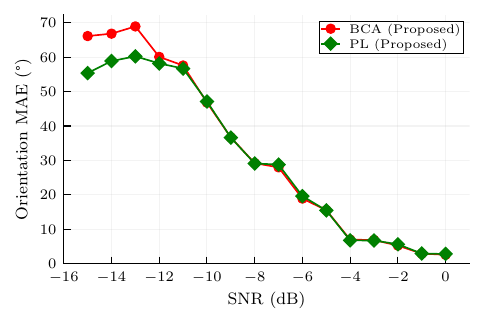}}%
    \subfloat[Beamwidth MAE, radar.]{\label{fig:radar_beta_mae}%
    \includegraphics[width=0.40\textwidth, height=\tempheight]{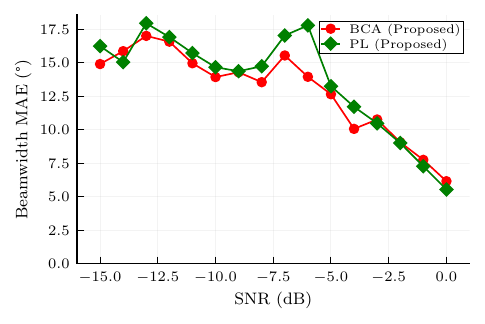}}%
    \caption{Beampattern parameter estimation error vs. SNR, aggregated over 250 Monte Carlo trials. Left: orientation $\hat{\phi}$. Right: beamwidth $\hat{\beta}$. The beamwidth is well identified in the radar scenario, whose spread formation samples the gain profile, and poorly identified in the communications scenario, where only a subset of receivers is illuminated.}
    \label{fig:beam_mae}
\end{figure*}
\end{revisedblock}

\subsection{Robustness to Beampattern Mismatch} \label{subsec:mismatch}
The matched-model results above use the same von Mises beampattern in simulator and estimator. To test robustness to departures from this template, we repeat both scenarios with the simulator radiating from a uniform-weight $M_e=4$-element ULA at spacing $\Delta_e=\lambda/2$, while the estimator still fits a von Mises model. The ULA array factor and its CRLB, denoted CRLB-ULA and plotted alongside the empirical curves, are derived in Section~S2 of the supplementary material; its half-power beamwidth is about $26\degree$ against the template's $30\degree$, with the mismatch concentrated in the sidelobes and tails.

The bottom row of Fig.~\ref{fig:err_vs_snr} shows that the profile likelihood retains its qualitative behaviour under mismatch and approaches CRLB-ULA at high SNR in both scenarios. At $0$~dB, the profile likelihood comes within $7\%$ of CRLB-ULA in the communications scenario and within $13\%$ in the radar scenario. The Newton refinement on the constrained Rayleigh quotient is important under mismatch: the WLS warm-start no longer corresponds to the maximum-likelihood inner solve when the data deviates from the von Mises template, and the Newton step is what closes the gap to CRLB-ULA.

The oracle-$\beta$ BCA baseline of \cite{youJointDirectEstimation2023} fares poorly under the communications mismatch, plateauing at roughly $160$~m at high SNR while the proposed profile likelihood tracks CRLB-ULA at the metre scale. \revised{The oracle-$\beta$ beamwidth is fixed at the nominal $30\degree$, the value a catalog prior would supply for this emitter, and deliberately not the true ULA-equivalent of $26\degree$. The baseline is therefore given a plausible prior rather than ground-truth, which is the intended comparison.} The catalog-prior assumption in \cite{youJointDirectEstimation2023} cannot recover the ULA sidelobe structure \revised{under this nominal beamwidth}, because the BCA inner loop cannot escape the misspecified parametric subspace. The profile likelihood sidesteps this by relaxing the parametric constraint on the beampattern in the inner solve.

\subsection{Beampattern Sensitivity} \label{subsec:beampattern_sensitivity}
\begin{figure*}[!t]
    \centering
    \subfloat[Communications.]{\label{fig:comms_crlb_sensitivity}%
    \includegraphics[width=0.48\textwidth]{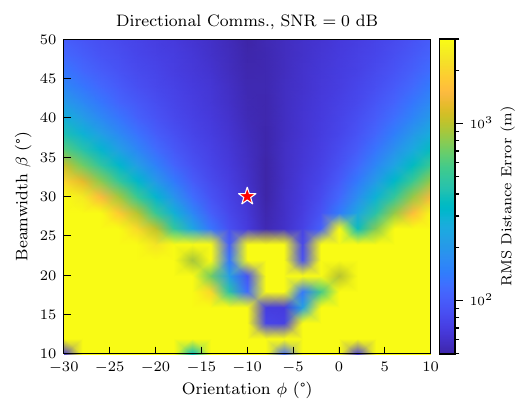}}
    \hfil
    \subfloat[Radar.]{\label{fig:radar_crlb_sensitivity}%
    \includegraphics[width=0.48\textwidth]{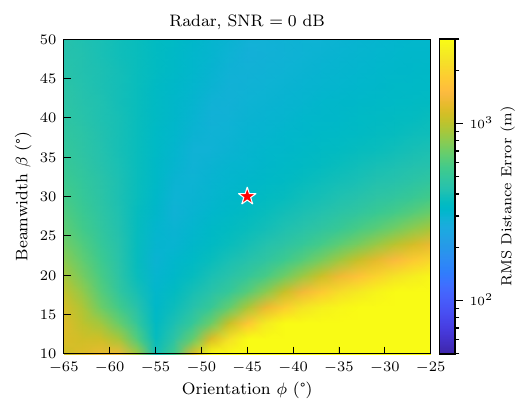}}
    \caption{Position CRLB over the beampattern-parameter plane $(\phi,\beta)$ at SNR $=0$~dB. The star marks the nominal operating point; the logarithmic colour scale saturates where the geometry becomes unobservable. Both scenarios form an accuracy basin that widens with $\beta$; the communications basin is narrow while the radar basin is broad and shallow.}
    \label{fig:crlb_sensitivity}
\end{figure*}

Fig.~\ref{fig:crlb_sensitivity} maps the position bound over the beampattern-parameter plane $(\phi,\beta)$ at $0$~dB, one panel per scenario. The bound is governed by which receivers fall on the gradient region of the main lobe: shifting $\phi$ moves the lobe across the formation, while widening $\beta$ trades gradient steepness against the number of illuminated receivers. \revised{Both geometries form an accuracy basin around their nominal parameters and improve as the beam widens, since a wider lobe brings more of the formation into the illuminated region; the radar bound falls from $501$~m to $300$~m as $\beta$ grows from $20\degree$ to $40\degree$ at nominal orientation. The orientation response is sharper and geometry-specific. The comms basin is narrow: the bound is minimized near $\phi=-10\degree$ and rises about fivefold, from $64$ to $311$~m, by $\phi=-20\degree$ as the lobe swings away from the informative corners, and it runs off scale for narrow beams, where the four-corner formation falls outside the main lobe. The radar basin is broad and shallow, minimized near $\phi=-50\degree$ and rising only from $340$ to $417$~m toward $\phi=-35\degree$ as the spread formation moves off the gradient region. Both reflect the mechanism of \cref{subsec:geometry}: the bound improves when more receivers occupy the illuminated gradient region of the main lobe. Because the position bound scales near-uniformly with noise power, the relative sensitivity shown is essentially independent of SNR.}

\begin{figure*}[!t]
    \centering
    \subfloat[Fixed $L=4$, varying SNR.]{\label{fig:placement_snr}
    \includegraphics[width=7cm]{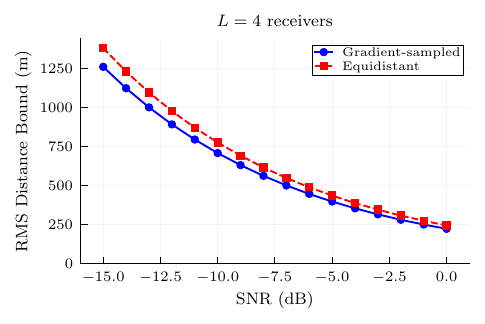}
    }
    \subfloat[Fixed SNR $=0$~dB, varying $L$.]{\label{fig:placement_l}
    \includegraphics[width=7cm]{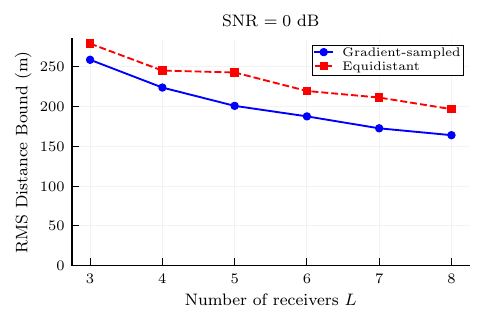}
    }
    \caption{\revised{RMS distance bound for gradient-sampled versus equidistant receiver placement in the radar scenario, at matched receiver count and angular sector. The gradient-sampled formation places receivers at the angles of steepest gain change; the equidistant formation spreads them uniformly, including near boresight where the RSS gradient vanishes. Each point averages $500$ signal realizations.}}
    \label{fig:crlb_placement}
\end{figure*}

\revised{The preceding observations concern where receivers sit relative to the beam rather than how many there are. Fig.~\ref{fig:crlb_placement} isolates this by comparing two placements of the same number of receivers within the same angular sector. The gradient-sampled formation splits receivers between the two angles of steepest gain change, at $\pm 17.8\degree$ off boresight for $\beta=30\degree$, just outside the $15\degree$ half-power edge where $|\partial g_l / \partial \theta_l^{(t)}|$ from (\ref{eq:gain_gradient}) peaks. The equidistant formation spreads them uniformly across the sector, so some sit near boresight where the gain is flat and the RSS gradient vanishes. The gradient-sampled formation attains a lower bound at every $L$, by $7\%$ to $18\%$, and the margin is independent of SNR since both formations scale identically with noise power. The advantage is largest at odd $L$, where the equidistant formation necessarily places one receiver exactly on boresight. Six gradient-sampled receivers outperform eight equidistant ones, so placement matters as much as receiver count.}

\begin{figure*}[!t]
    \centering
    \subfloat[Communications experiment.]{\label{fig:comms_cost_phi_beta}%
    \includegraphics[width=7cm]{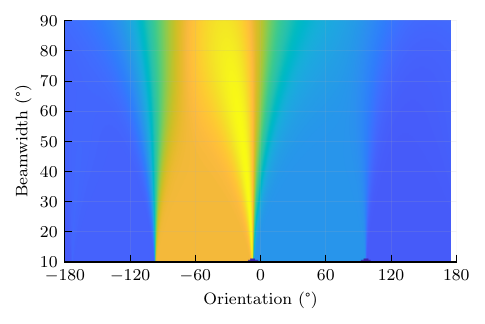}}
    \subfloat[Radar experiment.]{\label{fig:radar_cost_phi_beta}%
    \includegraphics[width=7cm]{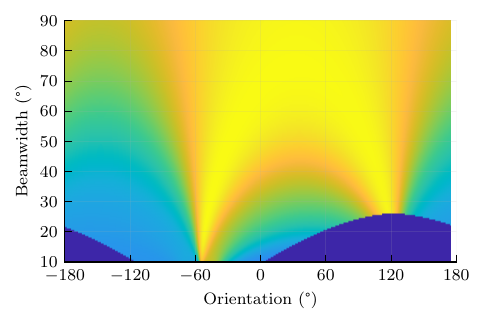}}
    \caption{Beampattern parameter cost surface $\tilde{Q}(\hat{\bm{p}}, \bm{\psi})$ at SNR $=0$~dB, evaluated at the true position.}
    \label{fig:cost_phi_beta}
\end{figure*}

Fig.~\ref{fig:cost_phi_beta} presents the beampattern parameter cost function surface $\tilde{Q}(\hat{\bm{p}}, \bm{\psi})$ evaluated at the true position for both scenarios at 0~dB SNR. For the communications experiment (Fig.~\ref{fig:comms_cost_phi_beta}), the cost function concentrates around the true orientation with a clear peak, confirming that the beampattern parameters are identifiable from the data. The radar experiment (Fig.~\ref{fig:radar_cost_phi_beta}) likewise produces a well-localized peak in orientation and a moderately broader ridge in beamwidth, reflecting that the spread formation samples enough of the gain profile to identify the orientation precisely while the beamwidth is constrained by the relative RSS at the boresight receivers versus \revised{those in the low-gain tail}.

\begin{revisedblock}
\subsection{Convergence Behaviour} \label{subsec:convergence}
The BCA cost rises monotonically, up to the grid quantization of the position block, and comes within $1\%$ of its converged value within two alternations at every SNR tested, taking a median of three at $0$~dB and four at $-10$~dB. Rapid cost convergence does not imply a correct position: at low SNR the iteration can settle in a poor local optimum (\cref{subsec:bca}), which is the source of the large BCA error in Fig.~\ref{fig:err_vs_snr}.

The safeguarded iteration is stable: the inner solve never diverges over all coarse-grid points and $100$ trials per SNR. Newton refinement averages $6.5$ iterations at $0$~dB, rising to $8.0$ at $-10$~dB, and the gradient fallback is exercised at a third to a half of grid points, so the inner problem is non-concave at points reached in normal operation and the safeguard is not a formality. Within a given basin the refinement barely moves the cost, by a median of at most $4.5\times10^{-3}$ in relative terms, so the warm start already sits at its local optimum and the method behaves as a profile likelihood despite the inexact inner maximization. Which basin is reached is set by the seed: the closed-form and enumerated seeds diverge on a fraction of trials, but the stationary points they reach are of equivalent localization accuracy (Section~S4), so we use the cheaper closed-form seed by default. The cost-versus-iteration trace, the full inner-solve statistics, and the seed comparison are given in Section~S4 of the supplementary material.

\subsection{Runtime} \label{subsec:runtime}
Table~\ref{tab:runtime} reports measured single-threaded times for one complete localization under the parameters of Table~\ref{tab:sim_params}, from a separate benchmark harness with all diagnostic instrumentation disabled. They bear out the complexity analysis of \cref{subsec:complexity}. BCA is the cheapest. The profile likelihood, seeded in closed form, runs in $164$~ms and $264$~ms, below the exhaustive search even at the shared $5\degree/2.5\degree$ beampattern grid ($301$~ms and $769$~ms).

The gap widens at matched accuracy. The exhaustive search resolves $\bm{\psi}$ only to its grid, whereas the Newton step resolves it continuously, so matching the profile likelihood's precision by exhaustive search needs a much finer grid; since (\ref{eq:cost_exhaustive}) is linear in $|\mathcal{G}_{\bm{\psi}}|$ the cost grows with it, reaching $3.13$~s and $8.73$~s at $1\degree$, more than an order of magnitude above the profile likelihood. Seeding by grid enumeration instead of the closed-form fit reaches the same accuracy (Section~S4) but pays for the $(\phi, \beta)$ enumeration, and is retained only as a fallback for ill-conditioned geometries.

The profile likelihood is therefore not the cheapest, but its case rests on accuracy at low SNR (\cref{subsec:localisation_performance}), on resolving $\bm{\psi}$ without a fine beampattern grid, and on structure: every position is evaluated independently, so the outer loop is embarrassingly parallel where the BCA alternation is sequential, which the single-threaded times in Table~\ref{tab:runtime} understate.

\begin{table}[!t]
    \centering
    \caption{Measured single-threaded runtime for one complete localization, mean over $10$ runs. Exhaustive is the four-dimensional joint search over $(\bm{p}, \bm{\psi})$.}
    \label{tab:runtime}
    \small
    \begin{tabular}{@{}lcc@{}}
        \hline
        Method & Communications & Radar \\
        \hline
        AOA--TDOA \cite{weissDirectPositionDetermination2004} & $16.0$~ms & $47.4$~ms \\
        BCA (Algorithm~\ref{alg:estimation}) & $47.2$~ms & $152.9$~ms \\
        Profile likelihood (Algorithm~\ref{alg:profile_likelihood}) & $164.1$~ms & $264.5$~ms \\
        Exhaustive, $\bm{\psi}$ grid at $5\degree/2.5\degree$ & $300.7$~ms & $769.2$~ms \\
        Exhaustive, $\bm{\psi}$ grid at $1\degree$ & $3.13$~s & $8.73$~s \\
        \hline
    \end{tabular}
\end{table}
\end{revisedblock}

\section{Conclusion} \label{sec:conclusion}
This article extended the DPD framework to directional emitters by jointly estimating position and beampattern parameters. Two algorithms were developed to handle the four-dimensional joint optimization: a BCA method that alternates between position and beampattern subproblems, and a profile likelihood (PL) method that concentrates the beampattern parameters out via an eigenvalue decomposition and a Newton-refined inner solve on the constrained Rayleigh quotient. The two differ structurally: BCA enforces the von Mises constraint throughout but is sensitive to its phase-only initialization and falls into poor local optima at low SNR, whereas the profile likelihood recovers the unconstrained amplitude pattern from a single eigendecomposition before fitting the model, trading an exact inner maximization for a well-conditioned, grid-independent inner solve that parallelizes across grid points.

Simulations on a communications and a radar scenario show the profile likelihood substantially outperforming both the AOA--TDOA baseline and BCA: it matches the CRLB above $-5$~dB in the communications scenario, with up to $50\%$ RMS error reduction over AOA--TDOA in the threshold regime, and tracks the CRLB across the operational range in the radar scenario while AOA--TDOA sits about a factor of two above the bound. The advantage is largest at low SNR, where the BCA initialization breaks down and AOA--TDOA lacks the RSS information to constrain position. An oracle-$\beta$ baseline mirroring \cite{youJointDirectEstimation2023} confirms that knowing the true beamwidth offers only marginal benefit, and the profile likelihood stays robust when the true pattern departs from the von Mises template.

Two structural properties of the cost function shape these results. Normalization by $\norm{\bm{d}(\bm{p}, \bm{\psi})}$ in (\ref{eq:qtildesol}) cancels absolute power, leaving localization invariant to transmit power and uniform system losses and elevating the beampattern parameters to the dominant sensitivities. The CRLB analysis further shows that the position bound is governed by which receivers occupy the gradient region of the main lobe, supporting deliberate spread of distributed formations across the beampattern profile rather than concentration near boresight; additional receivers help primarily by raising the probability that the formation samples this lobe-edge region.

\revised{The model assumes a common time reference across receivers, allowing the transmission time to be absorbed in (\ref{eq:freq_domain}). Residual clock offsets then act like ranging errors: a timing jitter of standard deviation $\sigma_\delta$ contributes on the order of $c\,\sigma_\delta$ to the position error. The beampattern estimation depends on the relative RSS rather than absolute timing, so these errors degrade the position while leaving the orientation and beamwidth essentially unaffected. Mitigation relies on disciplined clocks or two-way time transfer, or on TDOA formulations that estimate the offsets jointly with position \cite{xuHighAccuracyTDOALocalization2013}.}

\revised{The chief simplification is the von Mises beampattern itself: two parameters describe a single symmetric main lobe with a real-valued gain, leaving sidelobes, asymmetry, and any angle-dependent phase outside the model. The mismatch study of \cref{subsec:mismatch} shows the method tolerates a departure from this template, the simulator radiating from a uniform linear array, but the evaluation is still entirely synthetic. Validation against measured emitter patterns and field-collected data, where these effects are present, would be more convincing and is an important next step.}

The treatment here is restricted to a single stationary emitter observed in one snapshot. Future work will extend the framework to multi-emitter scenarios, multi-snapshot operation with moving platforms (where varying transmit angles provide additional angular diversity), and alternative parametric beampattern models.

\printbibliography

\begin{IEEEbiography}[{\includegraphics[width=1in,height=1.25in]{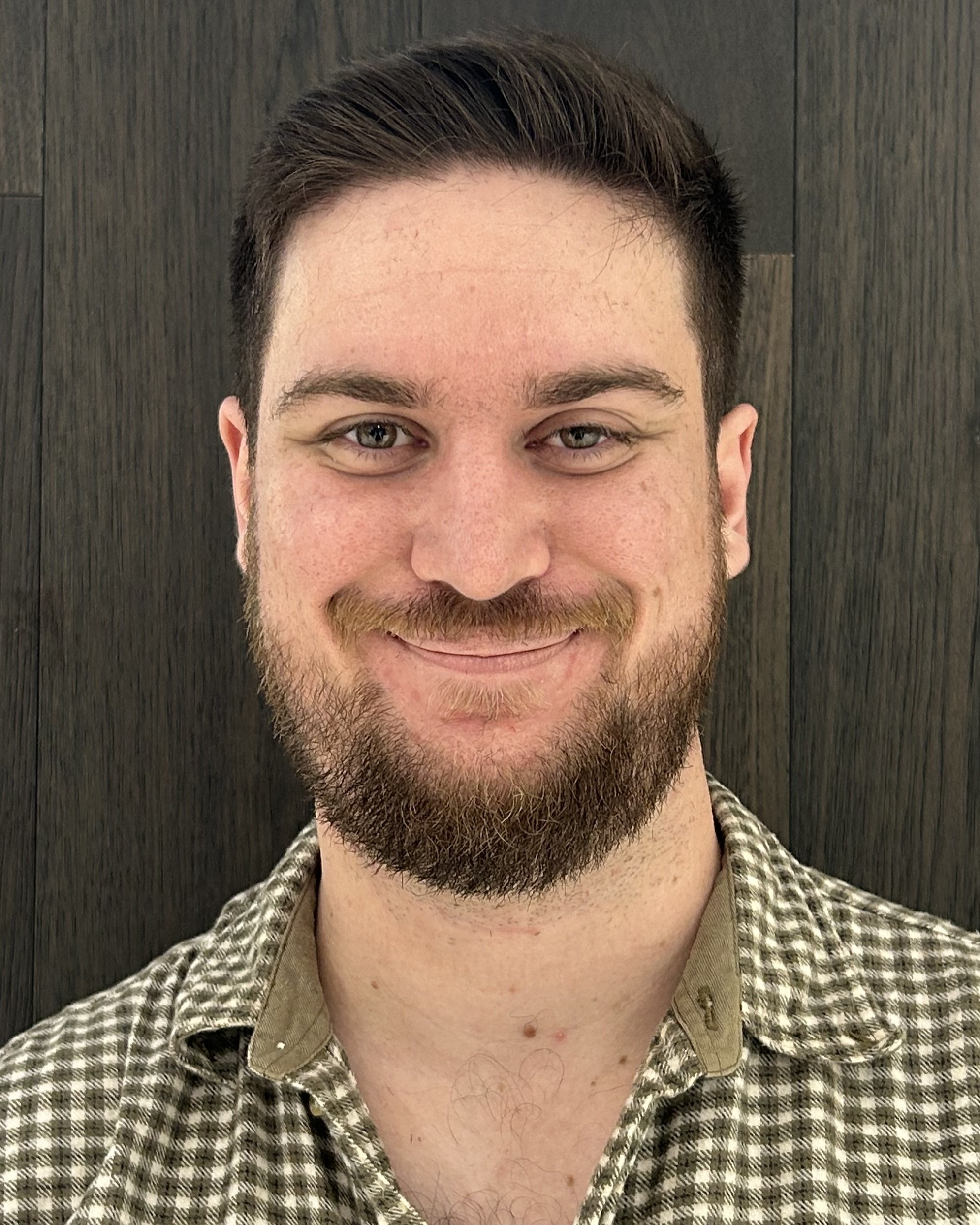}}]{Fraser Williams} received the B.Eng. in electrical engineering and B.Sc. in physics from the Queensland University of Technology, Australia, in 2021, and is pursuing the Ph.D. in electrical engineering at the same university.

His research interests include statistical signal processing, array processing, and radio frequency signal processing.
\end{IEEEbiography}%

\begin{IEEEbiography}[{\includegraphics[width=1in]{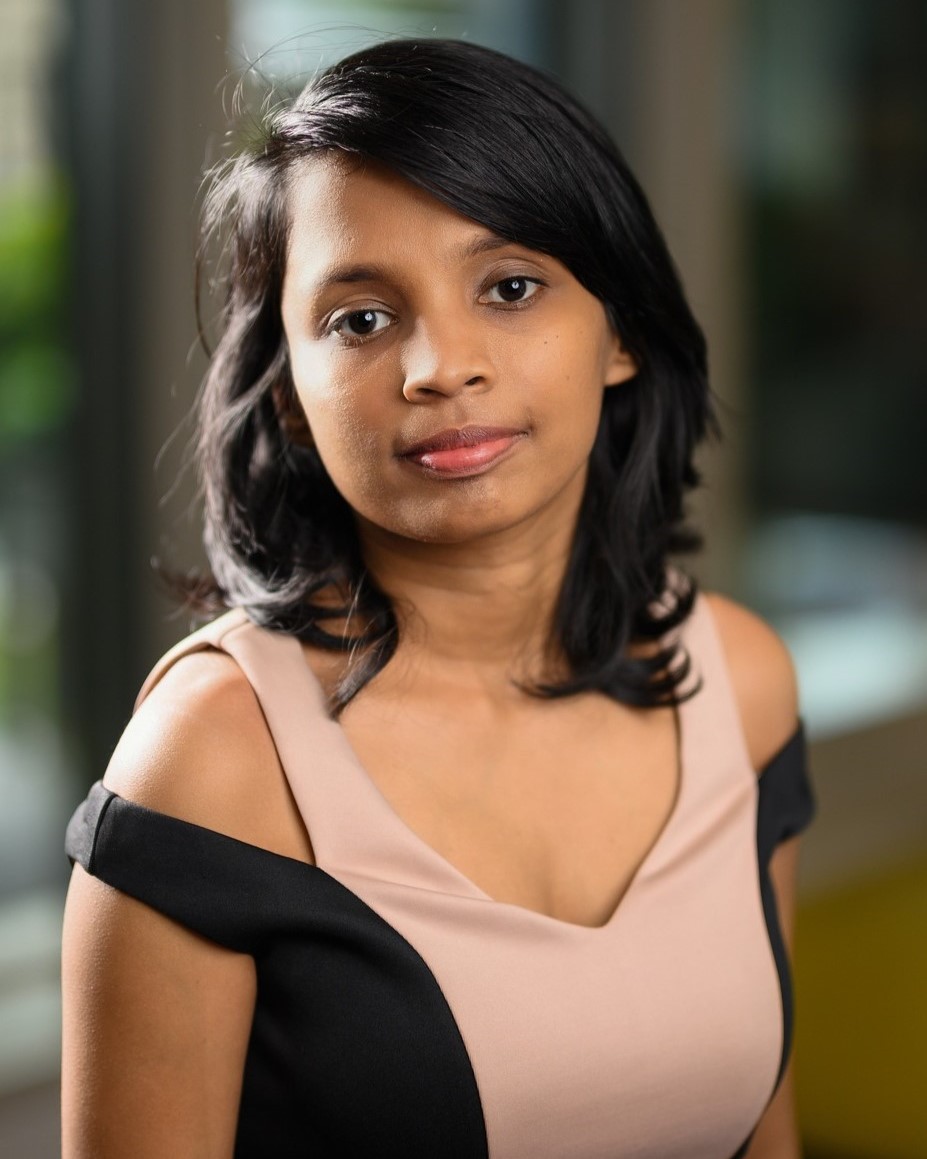}}]{Akila Pemasiri} received her B.Sc. degree in Computer Science and Engineering from the University of Moratuwa, Sri Lanka, and her Ph.D. from the Queensland University of Technology (QUT), Australia, in 2021. She is currently a Postdoctoral Research Fellow in the Signal Processing, Artificial Intelligence, and Vision Technologies group at QUT. Her research interests include applying machine learning principles to solve complex problems across various domains, including communications, defense, medical and healthcare, sports, and infrastructure.
\end{IEEEbiography}%

\begin{IEEEbiography}
[{\includegraphics[width=1in,height=1.25in]{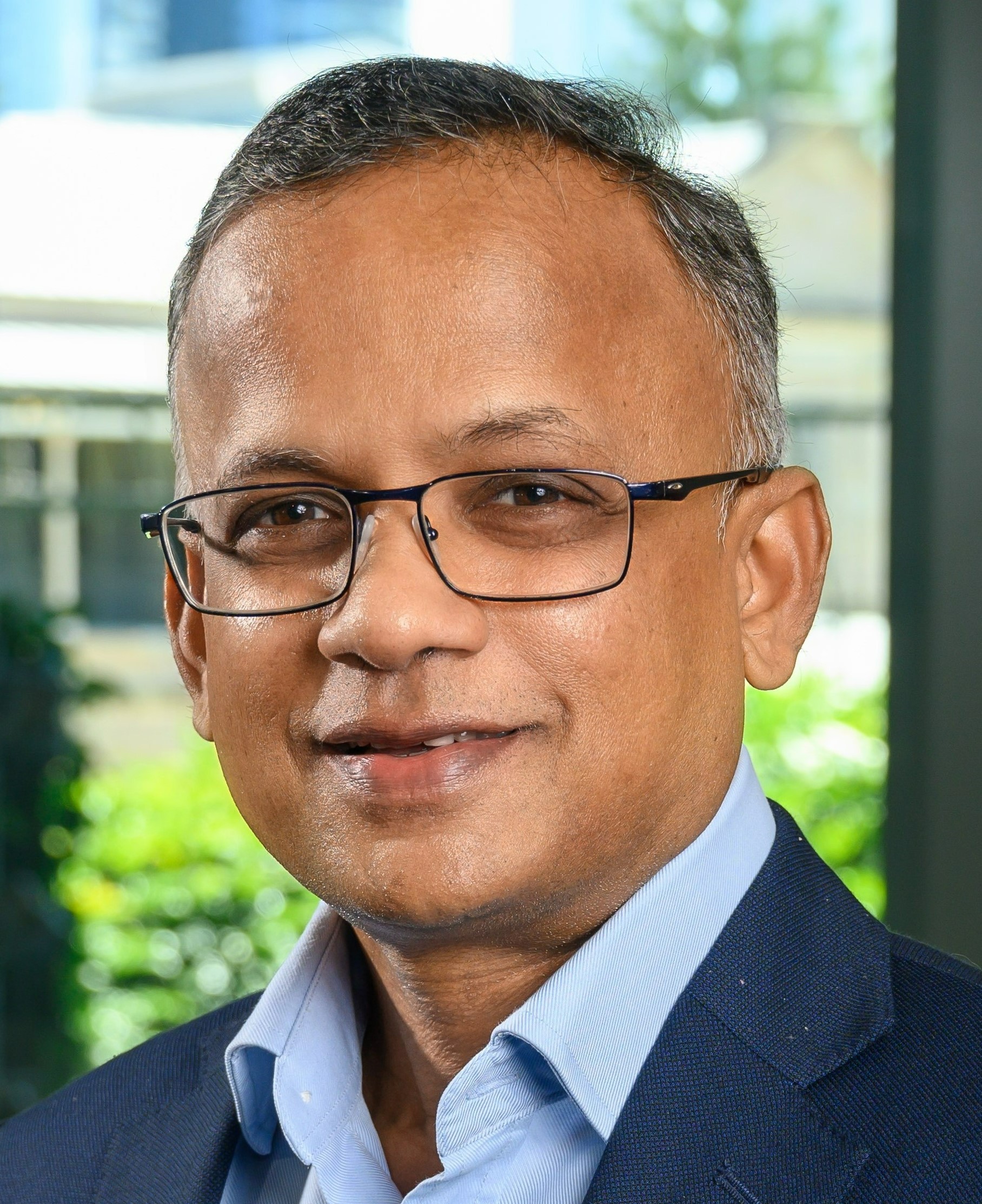}}]{Dhammika Jayalath}{\space}(Senior Member, IEEE) received the B.Sc. degree in electronics and telecommunications engineering from the University of Moratuwa, Sri Lanka, the M.Eng. degree in telecommunications from the Asian Institute of Technology, Thailand, and the Ph.D. degree in wireless communications from Monash University, Australia, in 2002. He was a fellow with the Australian National University and a Senior Researcher with the National ICT Australia. He has been an Academician with the Science and Engineering Faculty, Queensland University of Technology, since 2007. His research interests include the general areas of communications and signal processing. He has published significantly in these areas. His current research interests include applying machine learning principles in communication systems, physical layer security, and signal design for robust communications.
\end{IEEEbiography}%

\begin{IEEEbiography}[{\includegraphics[width=1in,height=1.25in]{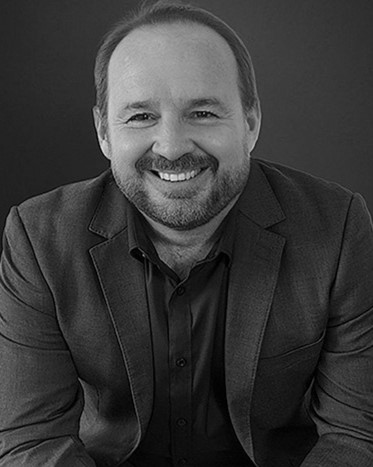}}]{Terrence Martin} is a co-founder of Revolution Aerospace Pty Ltd. He is a former RAAF \& Army military aerospace engineering officer with a Ph.D. in Applied Signal Processing \& Machine Learning. He has accumulated significant experience across a 35-year career, with time on fast jets \& rotary-wing manned platforms, alongside an extensive array of work for RAAF, Navy \& Army on a variety of UAV platforms. Terry has a deep interest in UAV-related technologies, and in 2018 he was nominated by Engineers Australia as one of Australia's 30 most Innovative Engineers.
\end{IEEEbiography}%

\begin{IEEEbiography}[{\includegraphics[width=1in,height=1.25in]{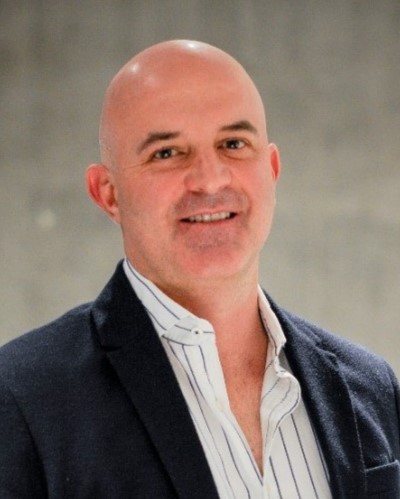}}]{Clinton Fookes}{\space}(Senior Member, IEEE) received the B.Eng. in Aerospace and Avionics, the MBA, and the Ph.D. degree in computer vision. He is currently the Associate Dean Research and a Professor of Vision and Signal Processing with the Faculty of Engineering at the Queensland University of Technology, Brisbane, Australia. His research interests include computer vision, machine learning, signal processing, and artificial intelligence. He serves on the editorial boards for IEEE TRANSACTIONS ON IMAGE PROCESSING and Pattern Recognition. He has previously served on the Editorial Board for IEEE TRANSACTIONS ON INFORMATION FORENSICS AND SECURITY. He is a Fellow of the International Association of Pattern Recognition and a Fellow of the Australian Academy of Technological Sciences and Engineering. He is a Senior Member of the IEEE and a multi-award-winning researcher including an Australian Institute of Policy and Science Young Tall Poppy, an Australian Museum Eureka Prize winner, Engineers Australia Engineering Excellence Award, Australian Defence Scientist of the Year, and a Senior Fulbright Scholar.
\end{IEEEbiography}%

\end{document}

%% file: preamble.tex
\usepackage{amsmath,amssymb,amsfonts}
\usepackage{graphicx}
\usepackage{textcomp}
\usepackage{tikz}
\usetikzlibrary{calc}
\usepackage{xcolor}
\usepackage{bm}
\usepackage{physics}
\usepackage[ruled,linesnumbered]{algorithm2e}
\usepackage{hyperref}
\usepackage{orcidlink}
\usepackage{gensymb}
\usepackage[font={sf,scriptsize},
            labelfont={bf},
            caption=false]
           {subfig}

\DeclareMathOperator*{\argmax}{\text{arg\,max}}
\DeclareMathOperator*{\argmin}{\text{arg\,min}}

\DeclareMathOperator{\arctantwo}{\text{tan}^{-1}}

\newcommand{\T}{\mathsf{T}}
\renewcommand{\H}{\mathsf{H}}
\DeclareMathSymbol{\shortminus}{\mathord}{AMSa}{"39}
\newcommand{\inv}{^{\shortminus 1}}

\newlength{\tempheight}
\newlength{\tempwidth}

\newcommand{\rowname}[1]%
{\rotatebox{90}{\makebox[\tempheight][c]{#1}}}

\newcommand{\columnname}[1]%
{\makebox[\tempwidth][c]{#1}}

\usepackage[capitalise,noabbrev]{cleveref}

\crefname{section}{Section}{Sections}
\Crefname{section}{Section}{Sections}
\crefname{subsection}{Section}{Sections}
\Crefname{subsection}{Section}{Sections}
\crefname{subsubsection}{Section}{Sections}
\Crefname{subsubsection}{Section}{Sections}

%% file: img/figures/experimental_setup.tex
\begin{figure}[!t]
    \centering
    \subfloat[Communication experiment.\label{fig:directive_comms_setup}]{
    \begin{tikzpicture}[x=0.7mm, y=0.7mm,font=\small]
        \draw[step=10, gray!30, ultra thin] (-40,-40) grid (40,40);
        \draw[->] (-40,0) -- (42,0) node[right] {$x$ (km)};
        \draw[->] (0,-40) -- (0,42) node[above] {$y$ (km)};

        \foreach \x/\y in {-25/-25, -25/25, 25/-25, 25/25} {
            \fill[blue] (\x,\y) circle (0.8);
            \draw[blue, thick] (\x-1,\y) -- (\x+1,\y);
            \draw[blue, thick] (\x,\y-1) -- (\x,\y+1);
        }

        \coordinate (emitter) at (6,6);
        \fill[red!10] (emitter) -- +(-10-15:50) arc (-25:5:50) -- cycle;
        \draw[red!40, thick] (emitter) -- +(-10-15:50);
        \draw[red!40, thick] (emitter) -- +(-10+15:50);
        \draw[red!40, thick] ([shift=(-10-15:50)]emitter) arc (-25:5:50);
        \fill[red] (emitter) circle (1);

        \node[above, red] at (emitter) {E};
        \node[right, red] at ($(emitter) + (18,-2)$) {$\phi=-10\degree$};
        \node[below, blue] at (-25,25) {$R_2$};
        \node[below, blue] at (-25,-25) {$R_1$};
        \node[below, blue] at (25,25) {$R_4$};
        \node[below, blue] at (25,-25) {$R_3$};
        \node[red!70] at ($ (emitter) + (-16:42) $) {$\beta=30\degree$};
    \end{tikzpicture}
    }
    \\
    \subfloat[Radar experiment.\label{fig:directive_radar_setup}]{
    \begin{tikzpicture}[x=0.7mm, y=0.7mm, font=\small]
        \draw[step=10, gray!30, ultra thin] (-40,-40) grid (40,40);
        \draw[->] (-40,0) -- (42,0) node[right] {$x$ (km)};
        \draw[->] (0,-40) -- (0,42) node[above] {$y$ (km)};

        \coordinate (emitter) at (-40,40);
        \fill[red!10] (emitter) -- +(-45-15:50) arc (-60:-30:50) -- cycle;
        \draw[red!40, thick] (emitter) -- +(-45-15:50);
        \draw[red!40, thick] (emitter) -- +(-45+15:50);
        \draw[red!40, thick] ([shift=(-45-15:50)]emitter) arc (-60:-30:50);
        \fill[red] (emitter) circle (1);

        \foreach \x/\y/\n/\pos in {20/-20/1/right, 45/-35/2/right, -20/-15/3/left, -5/-35/4/below} {
            \fill[blue] (\x,\y) circle (0.8);
            \draw[blue, thick] (\x-1,\y) -- (\x+1,\y);
            \draw[blue, thick] (\x,\y-1) -- (\x,\y+1);
            \node[blue, \pos, inner sep=2pt] at (\x,\y) {$R_\n$};
        }

        \node[above, red] at (emitter) {E};
        \node[red] at ($ (emitter) + (18,-2) $) {$\phi=-45\degree$};
        \node[red!70] at ($ (emitter) + (-38:42) $) {$\beta=30\degree$};
    \end{tikzpicture}
    }
    \caption{Experimental setup: emitter position, orientation, beamwidth, and receiver positions.}
    \label{fig:experimental_setup}
\end{figure}